\documentclass[seceq,preprint]{ptptex}

\usepackage{graphicx}

\newcommand{\bea}{\begin{eqnarray}}
\newcommand{\eea}{\end{eqnarray}}
\newcommand{\beq}{\begin{equation}}
\newcommand{\eeq}{\end{equation}}

\usepackage{epsfig}
\usepackage{amsmath}
\usepackage{amssymb}
\usepackage{amsthm}
\usepackage{float}
\usepackage{graphics}
\usepackage{color}

\preprintnumber[5cm]{
YITP-09-11\\
KU-PH-003\\
LA-UR 09-01034\\
KEK-CP-221\\
KUNS-2193\\
UTCCS-P-52}

\title{
A new scheme for the running coupling constant
in gauge theories using Wilson loops} 

\author{
Erek~Bilgici$^1$\footnote{E-mail: erek.bilgici@uni-graz.at},
Antonino~Flachi$^2$\footnote{E-mail: flachi@yukawa.kyoto-u.ac.jp},
Etsuko~Itou$^{3}$\footnote{E-mail: itou@yukawa.kyoto-u.ac.jp},
Masafumi~Kurachi$^{4}$\footnote{E-mail: kurachi@lanl.gov},
C.-J~David~Lin$^{5, 6}$\footnote{E-mail: dlin@mail.nctu.edu.tw},
Hideo~Matsufuru$^7$\footnote{E-mail: hideo.matsufuru@kek.jp},
Hiroshi~Ohki$^{2, 8}$\footnote{E-mail: ohki@yukawa.kyoto-u.ac.jp},
Tetsuya~Onogi$^2$\footnote{E-mail: onogi@yukawa.kyoto-u.ac.jp}
and~Takeshi~Yamazaki$^9$\footnote{E-mail: yamazaki@ccs.tsukuba.ac.jp}
}

\inst{
{\small
$^1$ Institut f\"{u}r Physik, Universit\"{a}t Graz, A-8010 Graz, Austria\\
$^2$ Yukawa Institute for Theoretical Physics, Kyoto University, Kyoto 606-8502,
Japan\\
$^3$ Academic Support Center, Kogakuin University, Nakanomachi Hachioji, 192-0015,
Japan\\
$^4$ Theoretical Division T-2, Los Alamos National laboratory, Los Alamos, NM 87544,
USA\\
$^5$ Institute of Physics, National Chiao-Tung University, Hsinchu 300, Taiwan\\
$^6$ Physics Division, National Centre for Theoretical Sciences, Hsinchu 300, Taiwan\\
$^7$ High Energy Accelerator Research Organization (KEK), Tsukuba 305-0801, Japan\\
$^8$ Department of Physics, Kyoto University, Kyoto 606-8501, Japan\\
$^9$ Center for Computational Sciences, 
University of Tsukuba, Tsukuba, Ibaraki 305-8577, Japan\\
}
}

\abst{%
We propose a new renormalization scheme of the running coupling constant
in general gauge theories using the Wilson loops.
The renormalized coupling constant is obtained from
the Creutz ratio in lattice simulations and the corresponding perturbative
coefficient at the leading order.
The latter can be calculated by adopting the zeta-function
resummation techniques.
We perform a benchmark test of our scheme in quenched QCD with
the plaquette gauge action.
The running of the coupling constant is determined by applying
the step-scaling procedure.
Using several methods to improve the statistical accuracy, we show
that the running coupling constant can be determined in a wide range
of energy scales with relatively small number of gauge configurations. 
}

\begin{document}

\maketitle

\section{Introduction}

One of the key subjects upon which recent attention has been focused
is the flavor dependence of $SU(N)$ Yang-Mills theories.
In particular, given a number of flavors $N_f$, the question 
is whether the theory has an (approximate) infrared fixed point.
This question is triggered by efforts to construct an alternative
mechanism of electroweak symmetry breaking, {\it via} assuming the existence of
a new, strongly interacting sector beyond the electroweak scale
\cite{techrev}.
The earliest model of this sort, the so-called technicolor\cite{tc1,tc2},
gives rise to a dynamical electroweak symmetry breaking by introducing
a QCD-like sector scaled up to some TeV.
While theoretically appealing, the simplest form of the technicolor
model and its variants with QCD-like dynamics are ruled out or
disfavored by electroweak precision measurements.
However, the possibility of such mechanism with 
a non-QCD-like theory~\cite{wtc1, wtc2, chipt1, chipt2, my, chipt3}
is still open, and may provide observable signatures 
at the LHC.  It is thus an important but 
challenging task to investigate the low-energy landscape
of spontaneously broken, strongly interacting gauge theories~\cite{bz}. 

Among the theoretical tools
at hand, the numerical approach to lattice gauge
theories has made it possible
to gain quantitative information about strong
dynamics of gauge theories.
The current understanding can be summarized as follows. 
A vector-like gauge theory, {\it e.g.} QCD, is known to exhibit confinement and
dynamical chiral symmetry breaking for small number of massless fermions,
$N_f$, in the fundamental representation of the gauge group.
When $N_f$ is just below the value $N_f^{af}$ at which the 
asymptotic freedom sets in,
the theory is conformal (unbroken chiral symmetry, no confinement) in the infrared.
Such a theory is believed to remain conformal down to 
some critical value $N_f^c$, where the coupling becomes strong enough and
the transition to the confined chirally broken phase occurs.
The range $N_f^c \leq N_f \leq N_f^{af}$ is called the
{\it conformal window}.

It is thus essential to investigate strongly interacting gauge
theories in a wide range of parameters, such as the number of colors,
the number of flavors, and the fermion representations~\cite{ogs}.
Several modern lattice studies in this research direction have recently
been performed~\cite{Catterall:2007yx, Appelquist:2007hu, 
DelDebbio:2008tv, DelDebbio:2008zf, 
DelDebbio:2008wb, Catterall:2008qk, DeGrand:2008kx, Shamir:2008pb, Fodor:2008hm,
Fodor:2008hn, Hietanen:2008mr, Appelquist:2009, fleming}.
In particular, the authors of Refs.~\cite{Appelquist:2007hu, Appelquist:2009}
performed calculation of the running coupling constant using the
Schr\"{o}dinger functional scheme, and found
evidence for an infrared fixed point in $SU(3)$ gauge theory with
$N_{f}=12$.
However, it is important to study the
running coupling constant in different renormalization schemes,
in order to conclude that the fixed point is not an artifact
due to a particular renormalization prescription but a physical one.
For this purpose, we propose a new renormalization scheme
which uses the Wilson loops as a key ingredient. 
Such a scheme is applicable to general gauge
theories as long as the Wilson loops can be defined,
and provides an efficient computational method
for lattice gauge theories.
Specifically, a renormalized amplitude is defined as the ratio among
the Wilson loops, namely the Creutz ratio, and its perturbative
counterpart. 
The former can be evaluated non-perturbatively by Monte Carlo
simulation, while the latter calculated analytically once
the underlying theory is specified.
By properly defining the non-perturbatively renormalized coupling
constant, its scale dependence is extracted using the step-scaling
procedure, {\it i.e.,} from the volume dependence of the 
coupling~\cite{Luscher:1991wu}.

Applying our scheme will provide not only an independent check on the extent
of the conformal window, but also several computational advantages.
The Creutz ratio can be obtained without $O(a)$ discretization errors,
provided these errors are absent in the lattice action.
This means our scheme is in principle free from any $O(a)$ systematic 
effect.
Furthermore, this scheme only involves simple gluonic observables, therefore
does not introduce any particular kinematical setup
which can deteriorate the discretization error or break chiral symmetry.
Therefore it can be applied to simulations with dynamical fermions
of any type, without restrictions on $N_f$.
For these features, this scheme may be an attractive alternative
to the Schr\"odinger functional scheme or the twisted Polyakov loop
scheme\cite{deDivitiis:1994yz,deDivitiis:1994yp,deDivitiis:1993hj}.

Before performing calculations for the 
gauge theories with dynamical fermions,
as a benchmark test, we apply this new scheme to 
the computation of
the running coupling constant in quenched lattice QCD.
The numerical calculation is performed using the plaquette gauge action
with periodic boundary conditions. 
These boundary conditions are chosen for simplicity.
Nevertheless, it results in effects of degenerate
vacua known as the ``toron'' \cite{Coste:1985mn}.
Our scheme can, however, be applied in principle to any choice of
boundary conditions, such as twisted boundary conditions, which ensure
no unwanted zero-mode contributions by inducing non-trivial background
configurations.
Adopting several methods to improve statistical accuracy, we can
determine the running of the coupling constant in a wide
range of energy scales with a relatively small number of gauge
configurations.

Another essential ingredient of our scheme is the perturbative calculation
of the renormalization constant.
This is performed analytically using zeta function resummation techniques,
which prove to be quite convenient.
First of all, zeta function techniques offer a natural method to study
the analyticity (and regularity) properties of the perturbative
counterpart of the Creutz ratio. 
In addition, some algebraic rearrangements of zeta functions,
originally due to Chowla and Selberg \cite{chse}, allow us to recast the expressions
in terms of analytic functions accompanied by some exponentially
converging series, whose evaluation is almost trivial and requires
little computer power. 
The zeta function methods we apply can be easily extended to any
boundary conditions and to the case of the Polyakov lines
\cite{forthcoming}.

This paper is organized as follows.
In the next section we give the definition of the new scheme.  
The perturbative calculation is illustrated in Sec.~3.
Section~4 is devoted to the details of our numerical simulations,
after brief introduction to the step-scaling procedure.
Section~5 contains discussion on the numerical results and comparison with
other results in the literature.
Finally, Sec.~6 summarizes our conclusions.
The paper contains two appendices where technical details
and simulation parameters are reported.
Preliminary results of this work have been presented in Ref.
\cite{Bilgici:2008mt}

\section{Wilson Loop Scheme}

In this section, we define a new renormalization scheme, the
`{\it Wilson loop scheme}'.
Let us consider an amplitude ${\cal A}$ whose tree-level 
approximation is
\begin{equation}
{\cal A}^{\rm tree} = k g_0^2~,
\label{eq:tree}
\end{equation}
where $g_0$ is the bare coupling constant, and $k$ is a coefficient of
proportionality that does not depend on $g_0$ and can be explicitly calculated
for a given underlying theory. 
With a non-perturbatively calculated amplitude ${\cal A}^{NP}$ at
the scale $\mu$,
the renormalization constant
$Z(\mu)\equiv {\cal A}^{NP}(\mu) / {\cal A}^{\rm tree}$
relates the renormalized coupling constant, $g(\mu)$, to the bare one,
leading to the relation
\begin{equation}
g^2(\mu) = \frac{{\cal A}^{\rm NP}(\mu)}{k}~.
\label{eq:g-mu}
\end{equation}
Although in particle physics an $S$-matrix element, {\it i.e.} a
scattering amplitude, is usually adopted as ${\cal A}$,
to define $g(\mu)$ one can equivalently use any physical quantity
that can be perturbatively expanded and is proportional to
$g_0^2$ at the tree-level.

\begin{figure}[t]
\begin{center}
  \includegraphics[height=5cm]{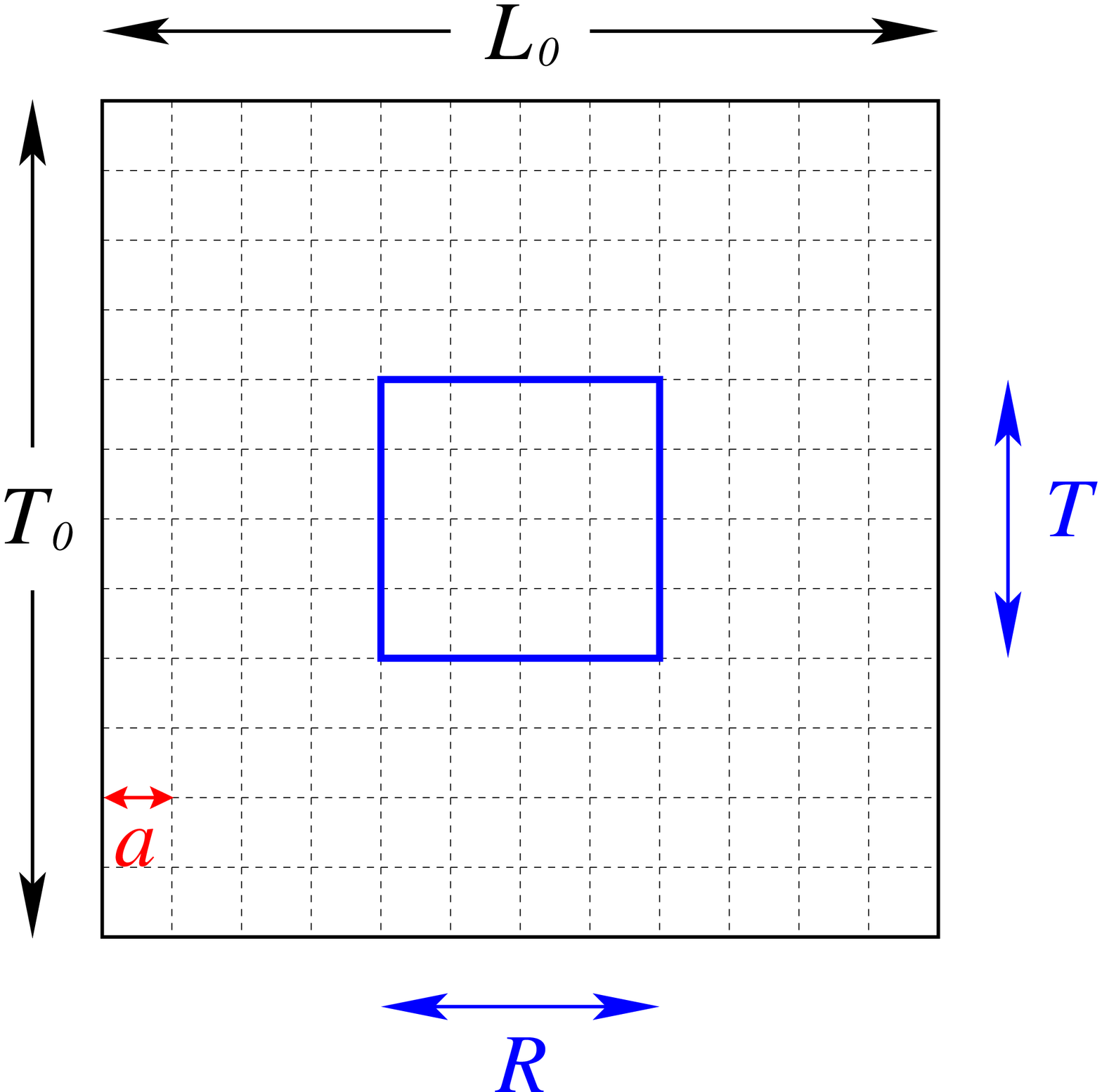} \vspace*{1cm}\\
\end{center}
\caption{Wilson loop defined on the latticized space-time box. $T_0$,  $L_0$ 
and $T$, $R$ represent the size of the box and the Wilson loop in the temporal 
and spatial directions, respectively; $a$ is the lattice spacing.}
\label{fig:Wloop}
\vskip 1cm
\begin{center}
  \includegraphics[height=2.1cm]{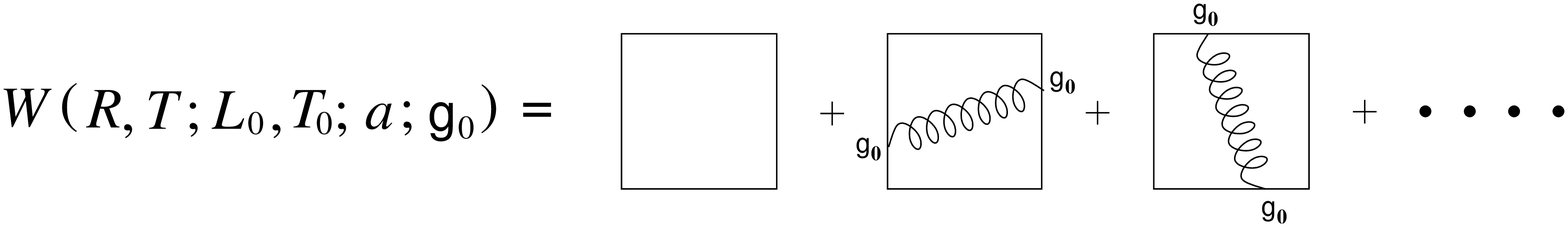}
\end{center}
   \caption{Schematic illustration of the perturbative expansion of the Wilson loop.}
\label{lpt}
\end{figure}

We define the Wilson loop scheme by taking the `amplitude' to be
\begin{equation}
{\cal A}_W (R; L_0; g_0) \equiv - R^2 \frac{\partial^2}{\partial R
 \partial T}\ln \left. \langle W(R, T; L_0, T_0) \rangle \right|_{T=R; T_{0}=L_{0}}, 
\end{equation}
where $W(R, T; L_0, T_0)$ is the Wilson loop with the temporal and spatial sizes
$T$ and $R$, on a lattice of the physical size $L_{0}^{3}\times T_{0}$.
In this work, we take $T_{0}$ to be the same as $L_{0}$, and drop it in the argument of the 
Wilson loop.  The scale $L_{0}$
will be identified as the renormalization scale later.
On a finite lattice, $W$, and thus ${\cal A}_W$, also depend on
the lattice spacing $a$ which is determined by the bare coupling $g_0$.
The dependence of ${\cal A}_W$ on $a$ is removed by taking the continuum
limit, $a\rightarrow 0$.
A pictorial definition of the Wilson loop is shown in Fig.~\ref{fig:Wloop}. 
Using lattice perturbation theory (see Fig.~\ref{lpt}), ${\cal A}_W$ can
be shown to be proportional to $g_0^2$ at the 
lowest order. Thus, once the value of $k$ is calculated,
relation (\ref{eq:g-mu}) leads to, after taking the continuum limit, 
a prescription to obtain the renormalized coupling:
\begin{equation}
 g^2\left(L_0, \frac{R}{L_0}\right) =
- {R^2\over k(R/L_0)} 
{\partial^2\over \partial R \partial T}\ln 
 \left. \langle W(R, T; L_0) \rangle^{\rm NP}\right|_{T=R} .
 \label{eq:g-L0}
\end{equation}
In the above expression we have made explicit that in the continuum limit $k$
is a {\it regular} function of $R/L_0$ only. This will be proved in
the next section. 
The remaining factor on the right hand side of Eq.~(\ref{eq:g-L0})
can be evaluated on the lattice as the Creutz ratio,
\begin{equation}
\chi(\hat{R}+1/2;  L_0/a)  = 
- \left. \ln 
\left(  
\frac{W(\hat{R}+1, \hat{T}+1;  L_0/a)\ W(\hat{R}, \hat{T};  L_0/a)}
{W(\hat{R}+1, \hat{T};  L_0/a)\ W(\hat{R}, \hat{T}+1;  L_0/a)}
\right) \right |_{\hat{T}=\hat{R}},
\end{equation}
where $\hat{T} \equiv T/a$ and $\hat{R} \equiv R/a$.
The value of $\chi$ is evaluated by a Monte Carlo simulation.

The renormalized coupling constant in the Wilson loop scheme can be written
as
\beq
g_{w}^2 \left(L_0, \frac{R+a/2}{L_0},\frac{a}{L_0} \right) = 
(\hat{R}+1/2)^2 \cdot \chi(\hat{R}+1/2;  L_0/a)/k .
\label{eq:def-g-wilson}
\eeq
The quantity $g_w^2$ depends on three different scales, $L_0$, $R$, 
and $a$; by taking the ratio to $L_0$, we use 
$r\equiv (R+a/2)/L_0$, $a/L_0$, and $L_0$ as the independent parameters.
Fixing $r$ to a specific value means fixing the renormalization scheme. 
The ratio $a/L_0$ specifies the discretization of the box, and can
be removed by taking the continuum limit, $a/L_0\rightarrow 0$.
After fixing the two dimensionless parameters $r$ and $a/L_0$, 
$g_w^2$ becomes a function of single scale, $L_0$.
In our scheme, following the step-scaling procedure, $L_0$ 
is identified as the scale at which the renormalized coupling is defined.

Ref.~\cite{Campostrini:1994fw} offers an alternative definition of the QCD running coupling related to Wilson loops and discusses its finite-size effects in $L_0/R$. 
Differently from us, their scheme defines the coupling constant in the $\hat{T} \rightarrow \infty$ limit, in which the renormalized coupling is related to the quark force.

There are several advantages in using the Wilson loop scheme.
An evident one is that our scheme does not contain $O(a)$ systematic effects
as long as they are absent in the lattice action.
This is because the Creutz ratio is free from $O(a)$ discretization errors,
due to the automatic $O(a)$ improvements of the heavy quark propagator after
the redefinition of the mass and the wave function \cite{Necco:2001xg}. 
This is in contrast to the case of the Schr\"odinger functional scheme, 
in which the boundary counter terms give rise to additional
$O(a)$ systematic errors.  Such a particular kinematical setup also breaks 
chiral symmetry.
Furthermore, this scheme only involves simple gluonic observables
and can be easily applied to the case with any type of dynamical
fermions without restriction to the number of flavors.


\section{Computation of $k$}
\label{creutzratio}

One of the indispensable ingredients of the scheme presented in the previous section
is the calculation of the coefficient $k$ in Eq.~(\ref{eq:tree}).
It can be generically split into two terms:
\bea
       k = k_0+k_1~,
\eea
where $k_0$ represents the zero-mode contribution, while $k_1$ can be expressed as 
\bea
k_1=-2\,R^2\, C_F\, \frac{\partial^2}{\partial R\, \partial T}
\left[\frac{4}{(2\pi)^4} \sum_{n} {}^{'}
\left(\frac{\sin\frac{\pi n_0 T}{L_0}}{n_0}\right)^2
\frac{e^{i\frac{2 \pi n_3 R}{L_0}}}{n^2}\right]_{T=R}~,
\label{eqk}
\eea
where the summation is taken over integer values
of $n_i$ ($i$=0,\dots,3) except for the case $n_0=n_1=n_2=n_3=0$
(indicated by the prime in the sum), 
and $n^2 \equiv n_0^2+n_1^2+n_2^2+n_3^2$.
The zero mode contribution depends on the boundary conditions.
In the following, we will concentrate on the case of periodic boundary
conditions. 
In this case, $k_0$ was initially calculated in Ref.~\cite{Coste:1985mn}.
For SU$(3)$ gauge group, $k_0$ is given by:
\bea
k_0=\frac{2}{3}\, C_F \left(\frac{R}{L_0}\right)^4~.
\label{zemo}
\eea

The scope of this section is to present a method to compute the quantity $k$.
The method we develop will be illustrated for the case of periodic boundary
conditions, but it can be applied, with minor changes, to the case of twisted
or mixed boundary conditions. As we have seen, the contribution from the zero
mode, $k_0$, can be separated from the rest and is obviously regular. 
Thus to compute (and to prove
the regularity of) $k$, we  only need to consider $k_1$. Our starting point
is the quantity:
\bea
S(T/L_0,R/L_0)&\equiv&
\sum_{n_0=-\infty}^\infty {\sin {2\pi T\over L_0} n_0 \over n_0}\nonumber\\
&&\times
\left[
2\sum_{n_1,n_2}
\sum_{n_3=1}^\infty {\cos{2\pi R\over L_0} n_3\over
n_0^2+n_1^2+n_2^2+n_3^2}+
\sum_{n_1,n_2}^{~~\prime}
{1\over n_0^2+n_1^2+n_2^2}
\right]
~.
\label{bas}
\eea
$k_1$ can be obtained from $S(T/L_0,R/L_0)$ {\it via}
$$
k_1= -{R^2 C_F \over 2\pi^3 L_0}{\partial S\over \partial R}(T/L_0,R/L_0)\nonumber~.
$$
Although it is not possible to find a closed form for $S(T/L_0,R/L_0)$ in terms of 
elementary functions, the use of zeta function resummation techniques and basic
analytic continuation 
allows us to recast $S(T/L_0,R/L_0)$ into the form of a practically computable
quantity, and to prove the regularity of $k$ through an explicit calculation.
The computation is carried out in a few steps.
The first is the evaluation of the sum over $n_3$ by using the Poisson summation formula. 
Then, the summation over $n_1$ and $n_2$, is written 
in terms of the Epstein zeta functions. After these steps, the expression
of $S(T/L_{0},R/L_{0})$
becomes compact.  However, without further rearrangements, it is of little practical use.
To this aim, it is convenient to rewrite the Epstein zeta 
functions using the Chowla-Selberg formula that renders the zeta functions into
the form of elementary analytic functions 
plus some rapidly converging series.  
The subsequent step is to analytically perform the integrals introduced when using
the Poisson summation formula, and 
finally perform the remaining summations numerically. 
Although the above procedure may seem involved, the actual implementation is
rather simple. The method has also the bonus of providing a proof of the 
regularity of the Creutz ratio, as we will explicitly show in the following.

The first step of our procedure is to employ the Poisson summation 
formula:
\bea
\sum_{n_3=1}^\infty f(n_3) = -{1\over 2}f(0)+\int_0^\infty dt f(t) 
+2\sum_{n=1}^\infty \int_0^\infty f(t) \cos(2\pi n t)dt~.
\label{apf}
\eea
A straightforward application of the above relation to the function 
$S(T/L_0,R/L_0)$ gives:
\bea
S(T/L_0,R/L_0) &=& 
2
\sum_{n_0=-\infty}^\infty {\sin {2\pi T\over L_0} n_0 \over n_0}
\sum_{m=-\infty}^\infty
\int_0^\infty 
\cos\left(2\pi (m+{R/L_0}) t\right)\zeta_t(s;~n_0) dt~,
\label{sab2}\eea
where we have used the standard definition of the generalized Epstein zeta function:
\bea
\zeta_t(s;n_0)\equiv
\sum_{n_1=-\infty}^\infty
\sum_{n_2=-\infty}^\infty
\left(n_0^2+n_1^2+n_2^2
+t^2\right)^{-s}~.
\eea
The parameter $s$ is a regulator, introduced to perform the necessary 
analytical continuations. The limit $s\rightarrow 1$ will be taken 
at the end of the calculation. 
It is interesting that the function $S(T/L_0,R/L_0)$ can be entirely
written in terms of the integral function
\bea
{\cal Z}(\Omega)&=&\int_0^\infty 
\cos\left(2\pi \Omega t\right)\zeta_t(s;n_0) dt~.
\label{ic}
\eea
Although compact, the result Eq.~(\ref{sab2}) requires further manipulation. 
A useful way to handle these functions is to make use of the Chowla-Selberg formula. 
Refs.~\cite{elizalde,kirsten} develop the appropriate formalism that allows us to
express $\zeta_t(s;n_0)$ as the sum of analytic functions plus a rapidly converging
series:
\bea
\zeta_t(s;n_0) &=& \pi {\Gamma(s-1)\over \Gamma(s)}|n_0^2+t^2|^{(1-s)}
+{2\pi \over \Gamma(s)}\sum_{p,q=-\infty}^{\infty~~\prime} 
\left[
\pi^2\left({p^2}+{q^2}\right)\right]^{-(1-s)/2}\nonumber\\
&\times& \left(n_0^2+t^2\right)^{(1-s)/2}K_{1-s}\left(2\pi \sqrt{n_0^2+t^2}
\sqrt{{p^2}+{q^2}}\right)~.
\label{chse}
\eea
The other tool is the following integral formula (see Ref.~\cite{gradryz}):
\bea
&&\int_0^\infty \cos \left(2\pi\Omega t\right)
\left(t^2+n_0^2\right)^{(1-s)/2}
K_{1-s}\left(2\pi (p^2+q^2)^{1/2}(t^2+n_0^2)^{1/2}\right)dt
\label{intform2}\\
&=& \sqrt{\pi\over 2} \left(2\pi\sqrt{p^2+q^2}\right)^{1-s}
n_0^{1/2+(1-s)} 
\left(4\pi^2\left(\Omega ^2+p^2+q^2\right)\right)^{\frac{s-1}{2}-\frac{1}{4}}
\nonumber\\
&&\mbox{ }\mbox{ }\mbox{ }
\times K_{(s-1)-1/2}\left(2\pi n_0\sqrt{(p^2+q^2)+\Omega^2}\right)~.\nonumber
\eea
The procedure is now straightforward and consists in using the relations
(\ref{chse}) and (\ref{intform2}) in Eq.~(\ref{sab2}). 
Some computations lead to
\bea
Z(\Omega)&=&
{\sqrt{\pi}\Gamma(2-s) \over \Gamma(s)}2
\Gamma(s-1)\cos\left(\pi((1-s)+1/2)\right)
\left({n_0\over \pi \Omega}\right)^{(1-s)+1/2}K_{(s-1)-1/2}\left(2\pi \Omega
n_0\right)\nonumber\\
&+&(2\pi)^{3/2}{2^{-s}\over \Gamma(s)} \sum_{p,q=-\infty}^{\infty~~\prime} 
n_0^{1/2+(1-s)}\left[4\pi^2\left({p^2}+{q^2} +\Omega^2\right)
\right]^{(s-1)/2-1/4}\nonumber\\
&\times& K_{(s-1)-1/2}
\left[
2\pi n_0 
\left({p^2}+{q^2} +\Omega^2\right)
\right]^{1/2}~.
\label{reg}
\eea
It can be easily checked that, in the above expression,
the limit $s\rightarrow 1$ can be taken safely giving
\bea
{\cal Z}(\Omega)
&=&{\pi \over 2|\Omega|} e^{-2\pi |\Omega| n_0}+{\pi \over 2}
\sum_{p,q=-\infty}^{\infty~~\prime} 
\left({p^2}+{q^2} +\Omega^2\right)^{-1/2} 
e^{-2\pi n_0 \sqrt{{p^2}+{q^2} +\Omega^2}}~.
\label{ii}
\eea
It is evident that the quantity ${\cal Z}(\Omega)$ is
regular. 
We can now substitute Eq.~(\ref{ii}) into Eq.~(\ref{bas}), 
and rearrange it as
\bea
S(T/L_0,R/L_0)&=&{4\pi^2 T\over L_0} \left[
S_0(R/L_0)+S_{1}(T/L_0,R/L_0)\right]~,
\label{resdw}
\eea
where we have separated the $n_{0}=0$ contribution from
the remaining part which is exponentially suppressed. 
This separation leads to the definition
\bea 
S_0(R/L_0) &=&A_1(R/L_0)+A_2(R/L_0)+A_3(R/L_0)~,\\
S_{1}(T/L_0,R/L_0) &=&B_1(T/L_0,R/L_0)+B_2(T/L_0,R/L_0)+B_3(T/L_0,R/L_0)~,
\eea
where
\bea
A_1(R/L_0)&\equiv& \sum_{m=-\infty}^{+\infty}{1\over 2|m+(R/L_0)|}~,\label{A1}\\
A_2(R/L_0)&\equiv& 2 \sum_{m=-\infty}^{+\infty}\sum_{p,q=1}^{\infty} 
{1\over \left({p^2}+{q^2} +|m+(R/L_0)|^2\right)^{1/2} }~,\label{A2}\\
A_3(R/L_0)&\equiv&
2\sum_{m=-\infty}^{+\infty}\sum_{p=1}^{\infty}
{1\over \left({p^2}+|m+(R/L_0)|^2\right)^{1/2}}~,\label{A3}\\
B_1(T/L_0,R/L_0)&\equiv& \sum_{n_0=1}^\infty{\sin {2\pi T\over L_0} n_0 \over \pi
n_0 T/L_0} \sum_{m=-\infty}^{+\infty}
{e^{-2\pi |m+(R/L_0)|n_0}\over 2|m+(R/L_0)|}~,\label{B1}\\
B_2(T/L_0,R/L_0)&\equiv& 2 \sum_{n_0=1}^\infty{\sin {2\pi T\over L_0}n_0 \over \pi
n_0 T/L_0} \sum_{m=-\infty}^{+\infty}
\sum_{p,q=1}^{\infty}{e^{-2\pi n_0 \sqrt{{p^2}+{q^2} +|m+(R/L_0)|^2}}\over 
\left({p^2}+{q^2} +|m+(R/L_0)|^2\right)^{1/2} }~,\label{B2}\\
B_3(T/L_0,R/L_0)&\equiv& 2 \sum_{n_0=1}^\infty{\sin {2\pi T\over L_0} n_0 \over \pi
n_0 T/L_0} \sum_{m=-\infty}^{+\infty}
\sum_{p=1}^{\infty} {e^{-2\pi n_0 \sqrt{{p^2}+|m+(R/L_0)|^2}}\over 
\left({p^2}+|m+(R/L_0)|^2\right)^{1/2}}~.
\label{B3}
\eea
Due to the exponential suppression, the terms $B_1$, $B_2$, and $B_3$, and thus
$S_{1}(T/L_0,R/L_0)$, are clearly regular. 
Therefore, to prove the regularity of $k$, we only have to show that $S_0(R/L_0)$ is
also regular. 
To show that the terms (\ref{A1}), (\ref{A2}), and (\ref{A3}) also lead
to a regular expression for $S_0(R/L_0)$ (and to compute them),
it requires further manipulations. 

The first term, (\ref{A1}), can be computed analytically:
\bea
A_1(R/L_0)={1\over 2}\left[ -{L_0\over R} -\psi(R/L_0) +\psi(-R/L_0)\right]~,
\label{resa1}
\eea
where $\psi(x)$ is the Euler psi function \cite{gradryz}. The remaining two
terms, $A_{2}$ and $A_{3}$, 
can be rearranged by performing first the summation over $m$, and
then by using the Chowla-Selberg formula.
We leave the details in Appendix~\ref{app:summation}, and present the results here:
\bea
A_2(R/L_0)&=&8 \sum _{j=1}^{\infty } \cos (2 j \pi  R/L_0) K_0\left(2 j \pi
 \sqrt{p^2+q^2}\right)+ \mbox{($R$-independent terms)}~,
\label{resa2}\\
A_3(R/L_0)&=&8 \sum _{q,j=1}^{\infty } \cos (2 j \pi  R/L_0)
K_0\left(2 j \pi q\right)
+ \mbox{($R$-independent terms)}~.
\label{resa3}
\eea
Written as above, it is a trivial matter to see that, after taking
the derivative of $A_1$, $A_2$,
and $A_3$ with respect to $R$, $S_{0}(R/L_0)$, and thus $k$, is nicely behaved due
to the exponential fall-off of the Kelvin functions $K_\nu(z)$.

\begin{figure}[]
\begin{center}
\unitlength=1mm
   \includegraphics[height=7cm]{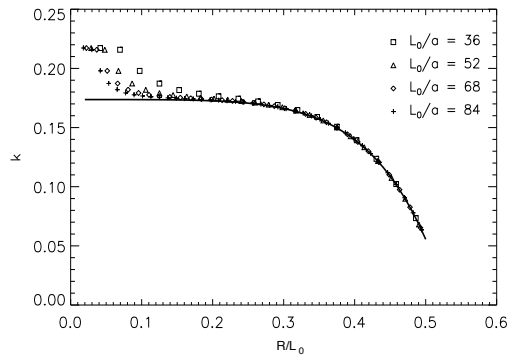}
\end{center}
\caption{The figure shows the dependence of $k$ on $R/L_0$.
The solid line represents $k$, in the continuum limit, according to the
analytical expressions given in the text. The squares are results obtained
using numerical lattice calculation for the sum in
Eq.~(\ref{eqk}). As can be seen from the figure, the
(continuum) limit of $k$ for $L_0/a \rightarrow \infty$ exists and is
finite. 
Also, the convergence of the lattice computation to the continuum value is
faster for larger values of $R/L_0$.} 
\label{fig:k}
\end{figure}

The last step of our procedure consists in evaluating the above expressions. 
The numerical computation of the sums in Eqs.~(\ref{B1}), (\ref{B2}), (\ref{B3}),
(\ref{resa2}), and (\ref{resa3}), does not present any problem due to the
exponential suppression.  Differentiating
with respect to $R$, substituting $T/L_0=R/L_0$, and combining the results
according to Eqs.~(\ref{eqk}), (\ref{zemo}), (\ref{bas}) lead to the 
result for $k$. Figure~\ref{fig:k} shows the dependence
of the function $k$ with respect to $R/L_0$. Table \ref{tab:k} provides some
indicative values for $k$.  

\begin{table}[]
       \centering
               \begin{tabular}{p{1cm}c|p{1cm}c}
                       $R/L_0$    & $k$                         & $\ R/L_0$    & $k$\\
                       \hline
                       $0.02$ & $0.17365\ $ & $\ 0.30$ & $0.16608$ \\
                       $0.10$ & $0.17360\ $ & $\ 0.35$ & $0.15694$ \\
                       $0.15$ & $0.17336\ $ & $\ 0.40$ & $0.13970$ \\
                       $0.20$ & $0.17259\ $ & $\ 0.45$ & $0.10885$ \\
                       $0.25$ & $0.17058\ $ & $\ 0.50$ & $0.05556$ \\ \\
               \end{tabular}
               \label{tab:k}
               \caption{Values for $k$ from the continuum calculation.}
\end{table}


\section{Numerical Simulation}

In this section, we will describe the details of our numerical simulations.
For later use, we define the coupling-squared, $\tilde{g}_w^2$, 
\beq
\tilde{g}_w^2\left(\beta,\, r,\, \frac{L_0}{a}\right) \equiv k g_w^2~,
\label{eq:tildeg}
\eeq
where $r\equiv(R+a/2)/L_0$. Note that we express $\tilde{g}_w^2$ as a function of
$\beta$, $r$, and $a/L_0$ 
instead of $L_0$, $r$ and $L_0/a$. The above redefinition is chosen for convenience,
since $\beta$, $r$, 
and $a/L_0$ are the actual input parameters for the simulations. 

\subsection{Step Scaling}
We begin by briefly reviewing the step-scaling procedure (see
Refs.~\cite{Luscher:1991wu, Luscher:1992an, Caracciolo:1994ed} for details),
that we use to evaluate the evolution of the running coupling in a wide
range of the energy scale on the lattice.

The first step is to fix a value for $r$, and find a set of parameters, $(\beta,
L_0/a)$, which produce the same value of $\tilde{g}_w^2$ for several different 
choices of $L_0/a$:
\beq
\left\{ \left( \beta^{(1)}_1, \left( L_0/a \right)^{(1)}_1 \right),\left(
\beta^{(1)}_2, \left( L_0/a \right)^{(1)}_2 \right),\cdots \right\}.
\eeq 
We achieve this by tuning the value of $\beta$ in such a way that the 
physical volume $L_0$ is fixed for different values of $L_0/a$. 
We denote this fixed physical volume for the starting point of the 
step-scaling procedure by $\tilde{L}_0$.

The next step is to vary the physical volume 
from $\tilde{L}_0$ to $s \tilde{L}_0$, which gives the evolution of 
the running coupling from the energy scale $\tilde{L}_0^{-1}$ to
$(s \tilde{L}_0)^{-1}$,
where $s$ is the scaling factor. This step can be performed by changing the
lattice size from $(L_0/a)^{(1)}$ to 
$s (L_0/a)^{(1)}$, leaving each value of $\beta^{(1)}$ unchanged.
Values of $g_w^2$ calculated with these new parameter sets should be considered as the 
coupling at the energy scale $(s \tilde{L}_0)^{-1}$ up to discretization errors,
and the extrapolation to the continuum limit can be taken
\beq
g^2_{R}\left( \frac{1}{s \tilde{L}_0}\right) \equiv 
\lim_{a \rightarrow 0}\left[
Z  \left(\frac{1}{s \tilde{L}_0}, \frac{a}{s \tilde{L}_0} \right)  g_{0}^2(a) 
\right] ,
\eeq
where $Z$ is the renormalization factor as defined below Eq.(\ref{eq:tree}).
The resultant value of the coupling, $g_R^2$, should be considered as the 
renormalized coupling at the energy scale $(s \tilde{L}_0)^{-1}$. This is the
way to 
obtain a single discrete step of evolution of the running coupling with scaling
factor $s$.

Next, we find a new parameter set for $(\beta^{(2)}, (L_0/a)^{(2)})$, which
reproduces the value of $g_w^2(1/s\tilde{L}_0)$ obtained in 
the previous step. 
Here, we chose the parameter set in such a way that the new lattice size
$(L_0/a)^{(2)}$ is equal to the original one, $(L_0/a)^{(1)}$.
From here, we can repeat exactly the same procedure described so far: we
calculate $g_w^2$ with the parameter set 
$(\beta^{(2)}, s(L_0/a)^{(1)})$.
By iterating this procedure $n$ times, we obtain the evolution 
of the running coupling from the energy scale $1/\tilde{L}_0$ to 
$(s^n\tilde{L}_0)^{-1}$.

\subsection{Simulation Parameters}

\begin{figure}[t]
\begin{center}
  \includegraphics[height=11cm]{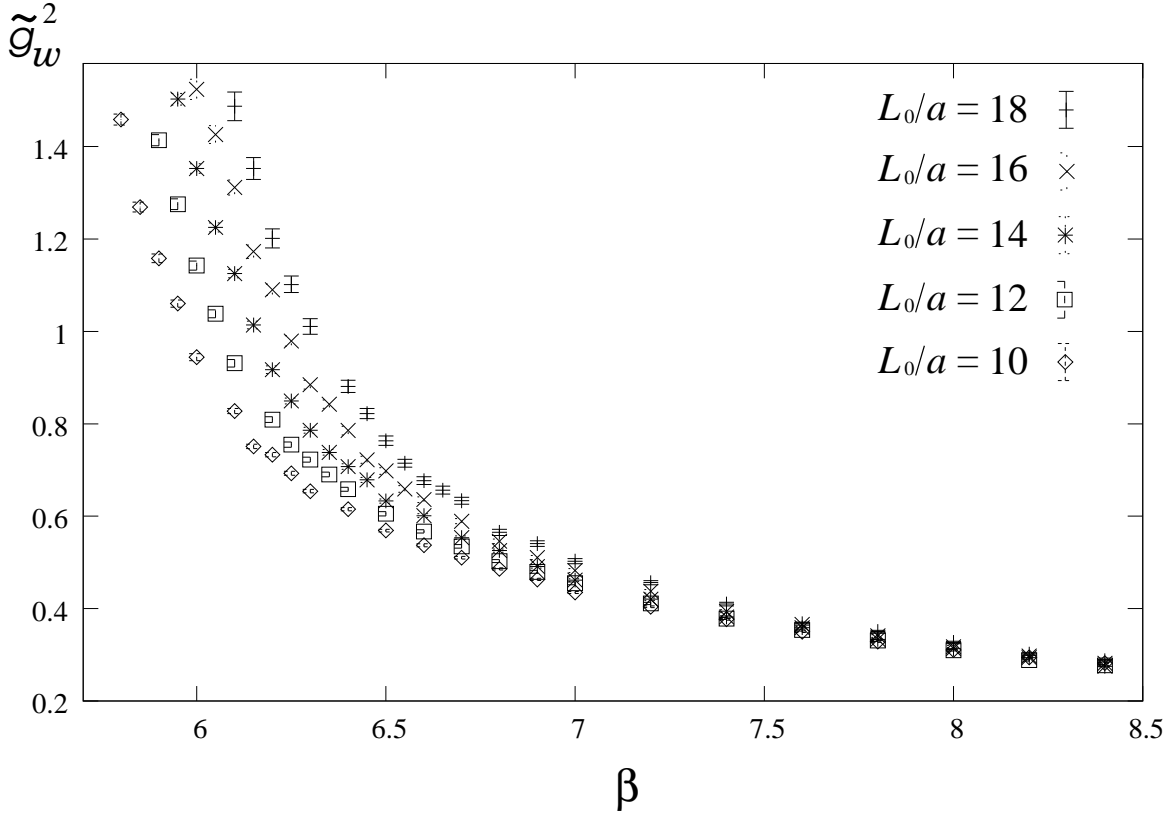}
\end{center}
\caption{$\tilde{g}_w^2$ obtained from simulations for different values 
of $L_0/a$ and $\beta$.}
\label{fig:g-beta-global}
\end{figure}

We use the standard Wilson plaquette gauge action defined on a 
four-dimensional Euclidean lattice with finite volume $L_0^4$. 
In this work,
we adopt untwisted periodic boundary conditions.  However,
it is straightforward to use other boundary conditions
({\it e.g.}, twisted boundary conditions) when necessary.
Gauge configurations are generated by using the pseudo-heatbath algorithm with 
over-relaxation, mixed in the ratio of $1$:$5$. In the remainder of this paper, 
we use the word ``a sweep" to refer to the combination of one pseudo-heatbath 
update sweep followed by five over-relaxation sweeps. In order to eliminate the 
influence of autocorrelation, we either take large enough number of sweeps 
between measurements, or adopt the method of binning with a large enough 
size of bin to estimate the statistical error reliably.
We perform the numerical simulations based on the step-scaling procedure 
explained in the previous section for a fixed value $r=0.3$. (The reason for 
this choice will be given in the next subsection.) 
We set the scaling parameter $s=1.5$ with five different starting
lattice sizes being $L_0/a = 10$, $12$, $14$, $16$ and $18$, 
which means lattice sizes after the scaling at
each step are $L_0/a = 15$, $18$, $21$, $24$ and $27$, respectively.
We take $\tilde{g}_w^2 = 0.2871$ 
(which corresponds to $g_w^2 = \frac{\tilde{g}_w^2}{k(r=0.3)} \simeq 1.728$) as 
the starting value of the first step of the step-scaling procedure.
Tunings of the values of $\beta$ (namely, finding values of $\beta$ which
satisfy
$\tilde{g}_w^2\left(\beta, r=0.3, \frac{L_0}{a}\right) = 0.2871$ for each 
$\frac{L_0}{a}=10$, $12$, $14$, $16$ and $18$ in 
the first step) are carried out by interpolating the data obtained from simulations 
for different values of $L_0/a$ and $\beta$ shown in Fig.~\ref{fig:g-beta-global}. 
Each data point in the figure is calculated from 
200 gauge configurations with 1000-sweep separation 
between configurations. 
Once we obtain values of $\beta$ which reproduce 
$\tilde{g}_w^2\left(\beta, r=0.3, \frac{L_0}{a}\right) = 0.2871$ for 
$\frac{L_0}{a}=10$, $12$, $14$, $16$ and $18$, we carry out simulation 
for $s=1.5$ step-scaling, namely simulations for $\frac{L_0}{a}=15$, $18$, 
$21$, $24$ and $27$ with the values of $\beta$ we tuned. These results are 
used to take the continuum limit, then the resultant value of $\tilde{g}_w^2$ 
becomes a starting value for the next step.
We iterate this procedure seven times. 
The combination of $L_0/a$ and $\beta$ used for the simulations are shown 
in Table~\ref{table:parameter-set}~\footnote{The values of $\beta$ in
Table II have numerical ambiguities coming from the statistical errors 
of the data used for the interpolation. These ambiguities propagate 
to the error of the physical scale at each step which we are trying to 
fix. However, it turned out that the effect of that fluctuation to the 
result of simulation for $s=1.5$
at each lattice size was negligibly small compared to the statistical
error of the simulation itself. Thus, we ignore those errors
and resultant fluctuation of the physical scale in the rest of our  
analysis.}.
\begin{table}[h]
\begin{center}
       \begin{tabular}{|c|c||c|c||c|c||c|c|}
       \hline
\multicolumn{2}{|c||}{Step $1$} & \multicolumn{2}{c||}{Step $2$} 
& \multicolumn{2}{c||}{Step $3$} & \multicolumn{2}{c|}{Step $4$}\\
\hline \hline
$L_0/a$ & $\beta$ & $L_0/a$ & $\beta$ & $L_0/a$ & $\beta$& $L_0/a$ & $\beta$\\
\hline
$15$ & $8.31$ & $15$ & $7.80$ & $15$ & $7.44$ & $15$ & $6.968$ \\
$18$ & $8.25$ & $18$ & $7.83$ & $18$ & $7.45$ & $18$ & $7.040$ \\
$21$ & $8.27$ & $21$ & $7.86$ & $21$ & $7.49$ & $21$ & $7.076$ \\
$24$ & $8.32$ & $24$ & $7.91$ & $24$ & $7.55$ & $24$ & $7.156$ \\
$27$ & $8.40$ & $27$ & $7.97$ & $27$ & $7.61$ & $27$ & $7.243$ \\
       \hline
       \end{tabular}
\vspace{4mm}
       \begin{tabular}{|c|c||c|c||c|c|}
       \hline
\multicolumn{2}{|c||}{Step $5$} & \multicolumn{2}{c||}{Step $6$} 
& \multicolumn{2}{c|}{Step $7$}\\
\hline \hline
$L_0/a$ & $\beta$ & $L_0/a$ & $\beta$ & $L_0/a$ & $\beta$\\
\hline
$15$ & $6.571$ & $15$ & $6.207$ & $15$ & $5.907$\\
$18$ & $6.656$ & $18$ & $6.303$ & $18$ & $6.000$\\
$21$ & $6.734$ & $21$ & $6.377$ & $21$ & $6.087$\\
$24$ & $6.797$ & $24$ & $6.463$ & $24$ & $6.170$\\
$27$ & $6.871$ & $27$ & $6.546$ & $27$ & $6.229$\\
       \hline
       \end{tabular}
\vspace{4mm}
\caption{Parameter sets, $L_0/a$ and $\beta$, used for the simulation.}
\label{table:parameter-set}
\end{center}
\end{table}


\subsection{Simulation Details}
\label{sec:SimDet}
There are several practical steps to calculate the quantity 
$\tilde{g}_w^2(\beta, r, L_0/a)$ from numerical simulations. 
Here we explain various technical details of our computations.

\begin{figure}[t]
\begin{center}
  \includegraphics[height=7.4cm]{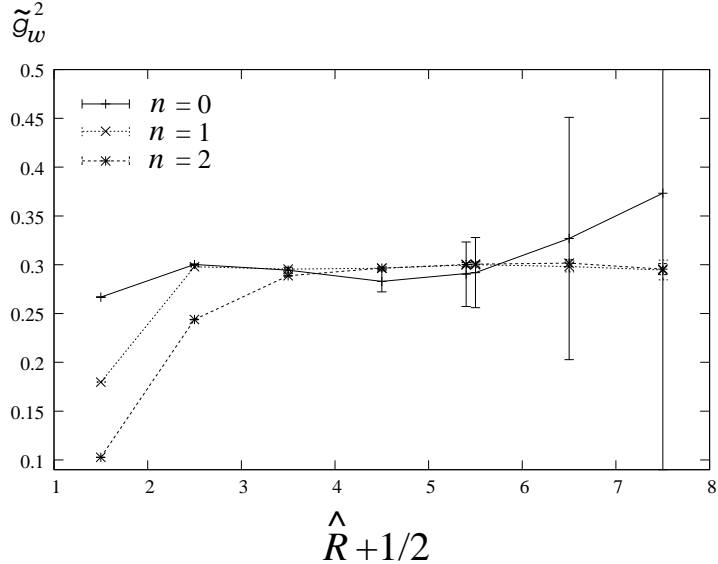}
\end{center}
\caption{The values of $\tilde{g}_w^2$ with statistical error for several values of 
$(\hat{R}+1/2)$ 
in the case of $\beta = 8.25$ and $L_0/a = 18$. Data connected by solid, dotted and 
dashed lines denote the data with $0$, $1$ and $2$ smearing levels, respectively.}
\label{fig:Smear-Plot}
\end{figure}

We use the APE smearing 
\cite{Albanese:1987ds} of link variables defined by the following equation;
\beq
U^{(n+1)}_{x,\mu}=Proj_{SU(3)} \left[ U^{(n)}_{x,\mu}+\frac{1}{c} \Sigma^{4}_{\mu
\ne \nu}U^{(n)}_{x,\nu} U^{(n)}_{x+\nu,\mu} U^{(n) \dag}_{x+\mu,\nu} \right],
\eeq
where $n$ and $c$ denote the
smearing level and the smearing parameter, respectively. 
The smearing is done for links in all four directions.
The result does not depend on the value of $c$ significantly, and we take $c=2.3$ 
in the present study.
Here, we need to find the optimal values of $r \equiv \frac{R+a/2}{L_0}$ and 
the smearing level $n$, by considering the following requirements. 
For better control of discretization error, it is preferable to choose a larger value 
of $r$. Meanwhile, for the purpose of reducing the statistical error, it is better to 
take a smaller value of $r$ and higher number of $n$. 
Fig.~\ref{fig:Smear-Plot} shows the  smearing-level dependence of 
$\tilde{g}_w^2$ in the case of $\beta=8.25$ and $L_0/a=18$ as an example.
From this figure, we find the statistical error is notably reduced even at
the smearing level one.
In order to avoid over-smearing, $n$ should be smaller than $\hat{R}/2$. 
This condition leads to the lower bound, $L_0/a > (4n+1)/(2r)$. 
We summarize the bound from this requirement
in Table~\ref{table:smearing-level}.
We observe 
(see Fig.~\ref{fig:Smear-Plot} for the example of the case $L_0/a=18$ at
$\beta = 8.25$)
that the data of $(\hat{R}+1/2)=1.5$ and $2.5$ in higher smearing level are not
reliable because of over-smearing.
By considering all the above requirements, we find that $(r,n)=(0.3, 1)$
is the optimal choice.

\begin{table}[h]

       \begin{center}
       \begin{tabular}{|c|c|c|c|}
       \hline
$n$     & $r=0.25$     & $r=0.30$     & $r=0.35$     \\
\hline \hline
$n=1$   & $L_0/a >10$  &$L_0/a >8.3$  &$L_0/a >7.1$  \\
$n=2$   & $L_0/a >18$  &$L_0/a >15$   &$L_0/a >12.8$ \\
$n=3$   & $L_0/a >26$  &$L_0/a >21.6$ &$L_0/a >18.5$ \\
       \hline
\end{tabular}
       \caption{The lower bound on $L_0/a$ to avoid over
smearing.}\label{table:smearing-level}
       \end{center}
       \end{table}

Once we fix the value of $r$ ($r=0.3$ in our current study), we need to 
estimate the value of $\tilde{g}_w^2$ for non-integer $\hat{R}$.
We interpolate the value of $\tilde{g}_w^2$ using a quadratic function:
\beq
f(\hat{R}+1/2)=c_0 +c_1 (\hat{R}+1/2) +c_2(\hat{R}+1/2)^2, 
\eeq
with interpolation ranges
for each lattice size listed in Table~\ref{table:fit-range}. 
An example is shown in 
Fig.~\ref{fig:Smear-Plot} where $(\hat{R}+1/2) = 5.4$ corresponds to the 
interpolation to $r=0.3$ in the case of $L_0/a=18$.

\begin{table}[h]

       \begin{center}
       \begin{tabular}{|c|c|c|c||c|c|c|c|}
       \hline
$L_0/a$ & $\hat{R}+1/2$ & $\hat{R}_{min}$ & $\hat{R}_{max}$ &
$L_0/a$ & $\hat{R}+1/2$ & $\hat{R}_{min}$ & $\hat{R}_{max}$     \\
\hline \hline
$10$   & $3.0$  &$2$  &$4$ & $18$   & $5.4$  &$4$  &$6$ \\
$12$   & $3.6$  &$2$  &$4$ & $21$   & $6.3$  &$5$  &$7$ \\
$14$   & $4.2$  &$2$  &$5$ & $24$   & $7.2$  &$5$  &$7$ \\
$16$   & $4.8$  &$3$  &$5$ & $27$   & $8.1$  &$6$  &$8$ \\
       \hline
\end{tabular}
       \caption{Ranges used to interpolate the value of $\tilde{g}_w^2$. 
       The column ``$\hat{R}+1/2$'' is the value that corresponds to  $r=0.3$.}
       \label{table:fit-range}
       \end{center}
       \end{table}

\begin{figure}[t]
\begin{center}
\includegraphics[height=6cm]{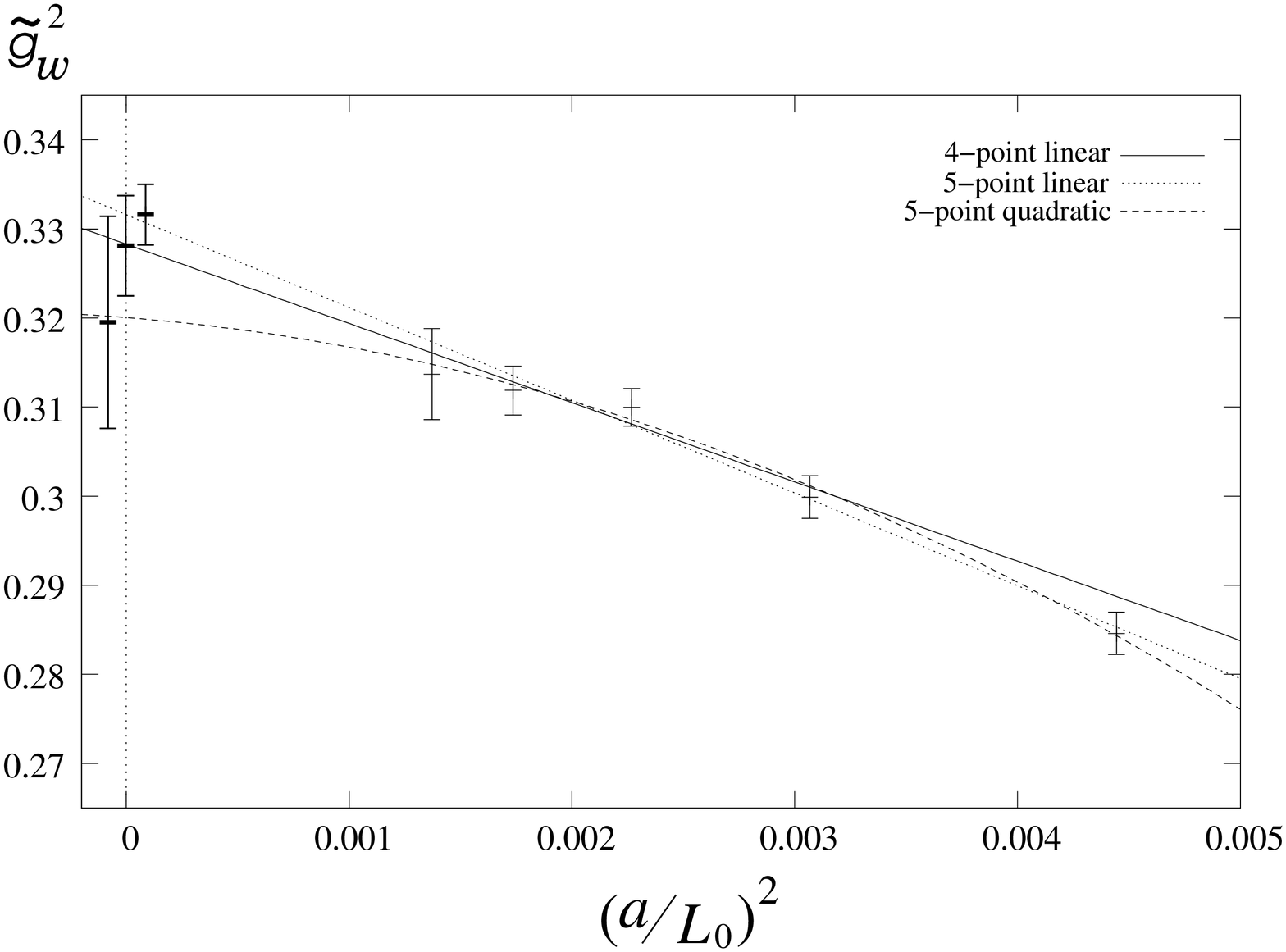}
\includegraphics[height=6cm]{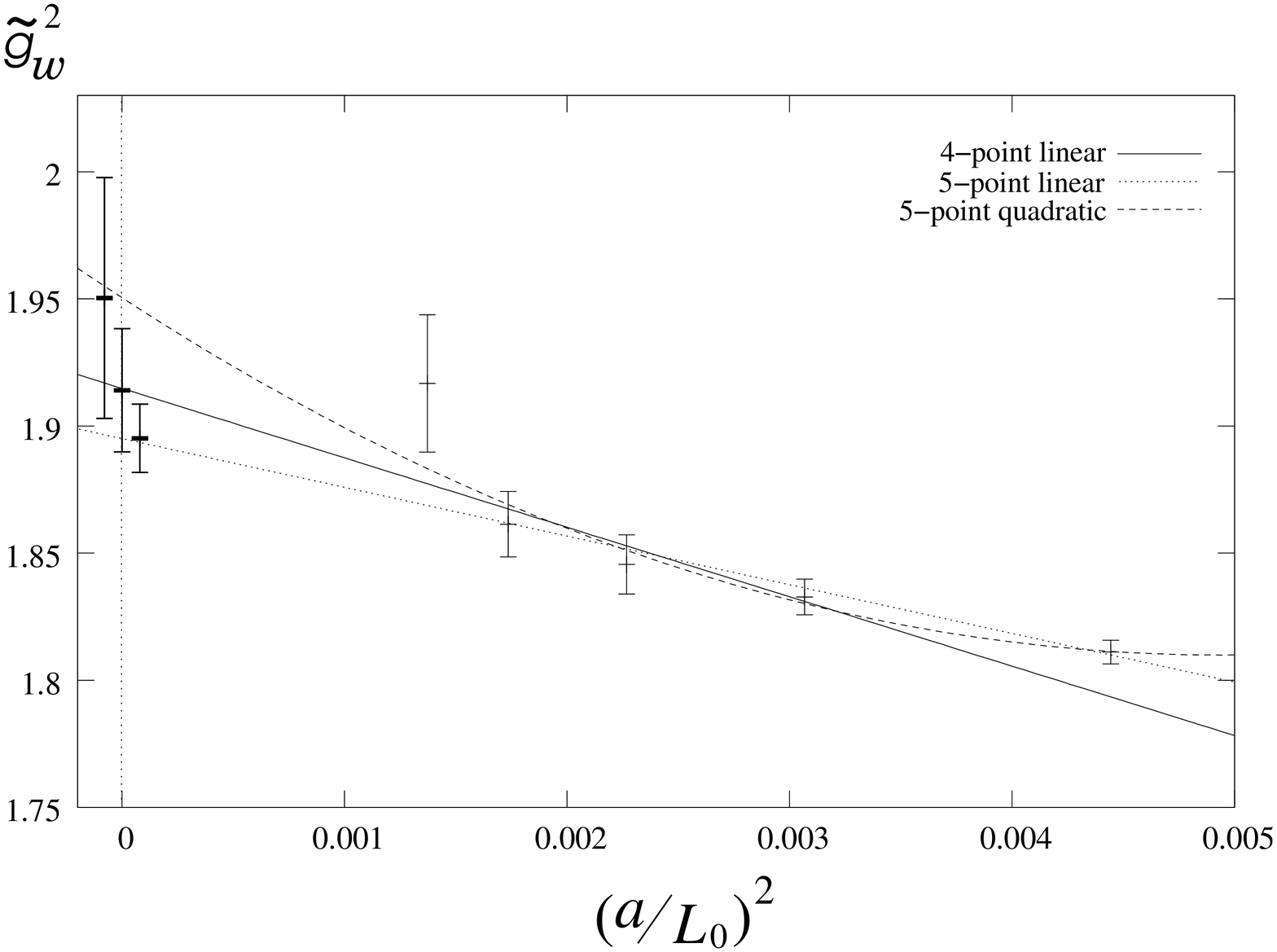}
\end{center}
\caption{The continuum limit of $\tilde{g}_w^2$. 
The left and right panels show Steps 1 and 7, respectively. 
}
\label{fig:Cont-Limit}
\end{figure}

The last step of the calculation is to take the continuum limit of $\tilde{g}_w^2$ 
from data obtained for different combinations of $\beta$ and $L_0/a$ 
listed in each column of Table~\ref{table:parameter-set}.
We show two example plots in Fig.~\ref{fig:Cont-Limit}, which are the continuum 
extrapolations for Step 1 and Step 7.
Since our Wilson loop scheme does not contain $O(a)$ systematic errors,
we extrapolate to the continuum limit 
using a fit function linear in $(a/L_0)^2$. Four data points
($L_0/a = 27$, $24$, $21$ and $18$) are used for this extrapolation
(shown as red lines in Fig.~\ref{fig:Cont-Limit}), and the 
resultant value is adopted as the central value of $\tilde{g}_w^2$ in the 
continuum limit.  We also take the continuum limit 
by using a fit function quadratic in $(a/L_0)^2$ with five data points 
($L_0/a = 27$, $24$, $21$, $18$ and $15$) (indicated by pink curves in 
Fig.~\ref{fig:Cont-Limit}), and the difference 
between the central values of two fits are adopted as 
the systematic error coming from possible higher order discretization effects. 
In Fig.~\ref{fig:Cont-Limit}, we have also plotted extrapolation by 
a linear function
with five points of data for comparison. In this figure, resultant values of 
the continuum limit  
obtained from different fit functions are plotted at $(a/L_0)^2=0$. 
(For better visibility, we slightly displaced the data obtained from 5-point 
quadratic and 5-point linear extrapolations.)  All the 
error bars shown in Fig.~\ref{fig:Cont-Limit} are statistical only.

\subsection{Numerical Results}
We now show the results of our simulations which were performed
using parameters in 
Table~\ref{table:parameter-set} with procedures explained in the previous section. 
Details of parameter choice and numerical results are summarized in 
Appendix~\ref{app:simulation_parameters}. 

\subsubsection{Running coupling}

\begin{figure}[t]
\begin{center}
\includegraphics[height=11cm]{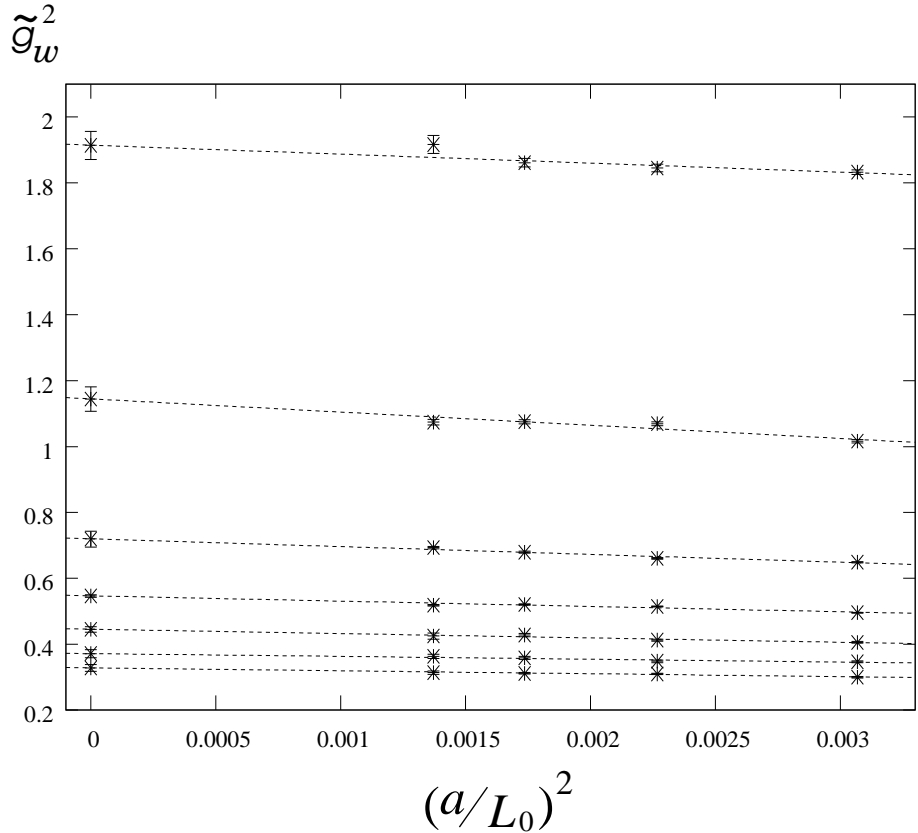}
\end{center}
\caption{Results of simulations and continuum limit of $g_w^2$ in 
Steps~$1 \sim 7$ (from bottom to top). 
}
\label{fig:contlimit_Step1-7}
\end{figure}
In Fig.~\ref{fig:contlimit_Step1-7}, we plot the resulting values of 
$\tilde{g}_w^2$ and their statistical errors for 
$L_0/a = 18$, $21$, $24$ and $27$ for Steps $1 \sim 7$. 
The continuum limit was taken in the way explained in the previous section, 
and both the statistical and systematic errors are estimated. In 
Fig.~\ref{fig:contlimit_Step1-7} the values of $\tilde{g}_w^2$ in the 
continuum limit are shown with statistical and systematic 
errors added in quadrature.

\begin{figure}[t]
\begin{center}
\includegraphics[height=9cm]{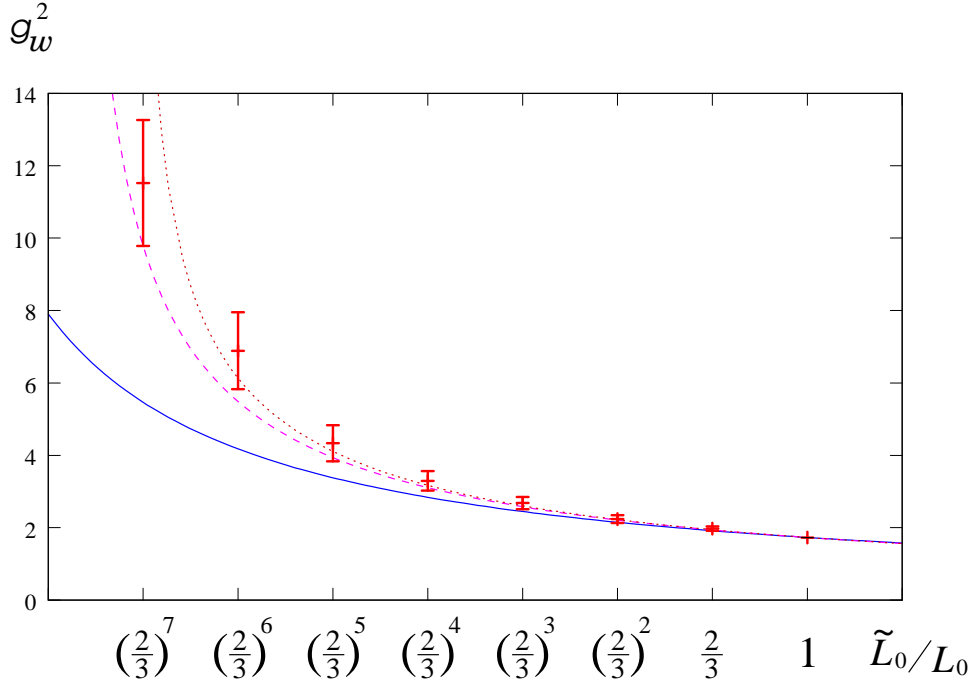}
\end{center}
\caption{Evolution of the running coupling constant in Wilson loop scheme, 
$g_w^2$, obtained from the step-scaling procedure. 
Horizontal axis shows the energy scale in units of 
$1/\tilde{L}_0$. Three curves, from bottom to top, 
show scheme-independent perturbative 
running couplings with one-loop and two-loop approximation, 
as well as that with three-loop approximation in the $\overline{{\mathrm{MS}}}$ scheme.}
\label{fig:run}
\end{figure}

The running coupling constant $g_w^2$ is extracted by dividing $\tilde{g}_w^2$ 
by $k(r=0.3)=0.1661$. The evolution of the running coupling constant 
is obtained by connecting the resultant values for Steps $1 \sim 7$ by 
assigning appropriate scales to these steps. We plot the results in 
Fig.~\ref{fig:run}.
We define the starting energy scale of Step $1$ as $1/\tilde{L}_0$, and 
the evolution of the running coupling constant is plotted as a function of 
energy in units of $1/\tilde{L}_0$.  
In this figure, errors are accumulated with the evolution of the running coupling 
appropriately\footnote{For the appropriate procedure of accumulating error, we need 
values of derivative of the step scaling function $\sigma(u)$. Here, we used the result 
of $u^5$ polynomial global fitting of $\sigma(u)$ to obtain approximate 
values of $\sigma'(u)$. See the next sub-subsection for detail.}
 in the same way as explained in Ref.~\cite{Capitani:1998mq} .
For comparison, we also plot scheme-independent perturbative 
running couplings with one-loop and two-loop approximation 
as well as three-loop approximation in the 
$\overline{{\mathrm{MS}}}$ scheme 
(from bottom to top).
In the high energy region, where the perturbative computation is reliable,
the Wilson loop scheme is consistent with the perturbation theory.
The figure also shows that our simulation is reaching deep into the 
low energy region, in which perturbative calculation is no longer reliable.

\subsubsection{Beta function}
\begin{figure}[t]
\begin{center}
   \includegraphics[height=8.7cm]{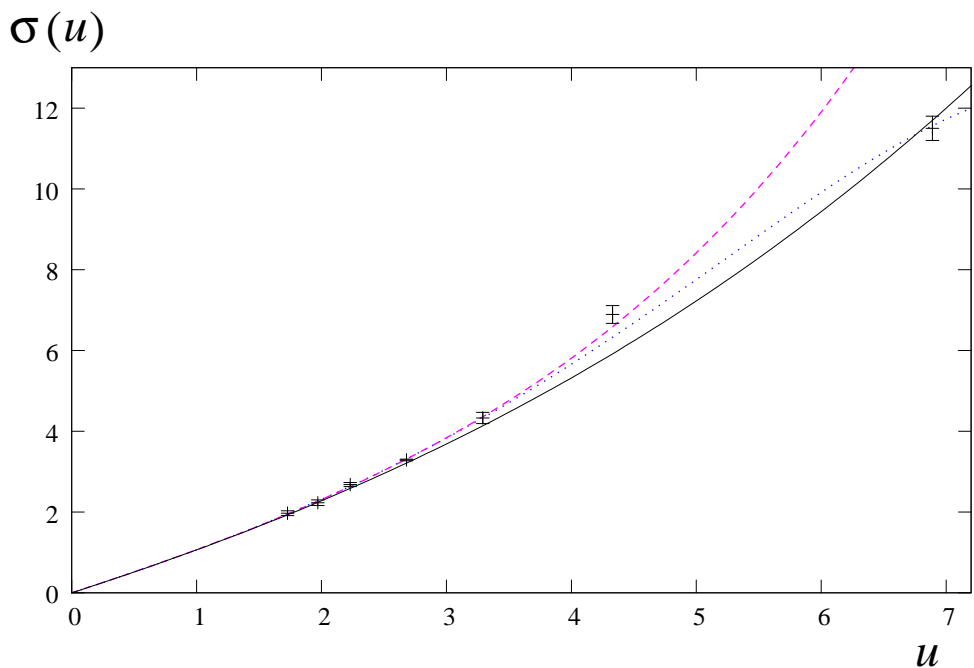}
\end{center}
\caption{Polynomial fit (dashed and dotted curves) of the data for the step 
scaling function $\sigma(u)$. Fit ansatz and resultant values of fitting 
parameters for two curves are explained in the text. The two-loop perturabative 
curve (solid) is also plotted for comparison.} 
\label{fig:step}
\end{figure}

From the results of the simulation, we can also extract the non-perturbative  
$\beta$ function by using the method explained in Ref.~\cite{DellaMorte:2004bc} . 
To this end, it is useful to define the step scaling function 
in the continuum limit as \cite{Capitani:1998mq}
\begin{equation}
\sigma(u) = g_w^2(sL),\ \  u\equiv g_w^2(L) .
\end{equation}
We list the simulation results for the step scaling function in Table \ref{tab:step-func}.
\begin{table}[h]
\begin{center}
\begin{tabular}{|c|c|}
\hline
$u$ & $\sigma(u)$ \\
\hline
1.728 & 1.97(6)  \\
1.97 & 2.23(7)  \\ 
2.23 & 2.68(5)  \\ 
2.68 & 3.29(2)  \\ 
3.29 & 4.33(14)  \\ 
4.33 & 6.89(22)  \\ 
6.89 & 11.5(3)  \\ 
\hline
\end{tabular}
\caption{
Simulation results for the step scaling function $\sigma(u)$. 
The values in parentheses represent total errors in units of 
the last digits.}
\label{tab:step-func}
\end{center}
\end{table}
In the week coupling region, $\sigma(u)$ can be perturbatively 
expanded as \cite{DellaMorte:2004bc}
\begin{equation}
 \sigma(u) = u + s_0 u^2 + s_1 u^3 + \cdots,
\end{equation}
with the coefficients
\begin{eqnarray}
s_0&=&2b_0 \ln s,\\
s_1&=&(2b_0\ln s)^2 + 2b_1 \ln s,
\end{eqnarray}
where $b_0$ and $b_1$ are the one-loop and the two-loop coefficients of the 
$\beta$ function, respectively. 
In Fig.~\ref{fig:step}, we plot the data listed in Table~\ref{tab:step-func} 
together with 
two-loop perturbative curve for $\sigma(u)$ (solid), as well as two  
curves which are the result of the following two kinds of polynomial fit. 
For the upper (dashed) curve, we fitted 
$u + s_0 u^2 + s_1 u^3 + s_2 u^4 $
to the data in the range of $u<5$, and obtained $s_2 = 0.0019(3)$.
For the middle (dotted) curve, we fitted 
$u + s_0 u^2 + s_1 u^3 + s_2 u^4 + s_3 u^5$
to all the data,  and obtained $s_2 = 0.0033(6)$ and $s_3=-0.00048(9)$.
As is expected, in the week coupling region, the data is well explained by 
the two-loop perturbative result, and 
polynomial functions fit well to the data also. 
Meanwhile, the figure clearly shows that neither two-loop perturbative curve, nor 
simple polynomial fits can explain the behavior of the data in larger values 
of $u$. This is nothing but an indication of the emergence of the non-perturbative effect. 

\begin{figure}[t]
\begin{center}
   \includegraphics[height=9cm]{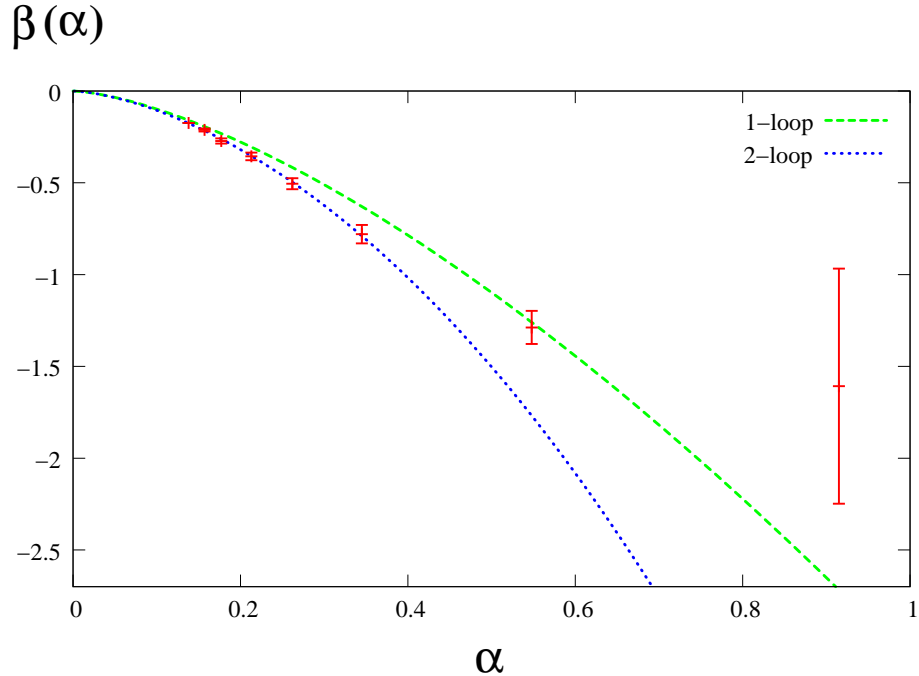}
\end{center}
\caption{Non-perturbative $\beta$ function in the Wilson-loop scheme.
One-loop (green dashed curve) and two-loop (blue dotted curve) perturbative 
$\beta$ functions are also plotted for comparison.
} 
\label{fig:beta}
\end{figure}

The formula to obtain non-perturbative $\beta$ function from the step scaling 
function was given in Ref.~\cite{DellaMorte:2004bc} as
\begin{equation}
  \beta \left(\sqrt{\sigma(u)}\right) = \beta \left(\sqrt{u}\right)
   \sqrt{\frac{u}{\sigma(u)}} \sigma'(u).
   \label{eq:beta}
\end{equation}
By applying this formula recursively, we obtained the discrete $\beta$ function 
as shown in Fig.~\ref{fig:beta}.
Here, while the values of the $\sigma(u)$ are directly given by the results of 
the simulation\footnote{Unlike was done in Ref.~\cite{DellaMorte:2004bc}, 
we do not have to carry out interpolation in $u$ since we have data 
exactly on the value of $u$ we need.}, we need to use the result of the 
fitting to obtain the values of $\sigma'(u)$. To obtain the latter, we 
adopted the $u^5$ global fitting explained above. As we noted, it 
fails to fit to the data in larger values of $u$, however, as can 
be seen from the figure, we can expect that it gives $numerically$ 
approximate values. Also, we used the value of perturbative two-loop 
approximation for the value of $\beta(u)$ at the smallest $u$ as an  
initial input value to use Eq.(\ref{eq:beta}) recursively. In the figure, 
we also plotted the one-loop and two-loop perturbative $\beta$ function 
for comparison. From this result, we confirmed again that for smaller coupling 
region, the data consistently reproduce the perturbative picture, 
while the non-perturbative effect begins to emerge for larger coupling region.

\section{Discussion}

We have hitherto concentrated on how the coupling constant
runs under the relative change of the scale without referring 
to its absolute value.  The absolute scale can be
estimated using the Sommer scale, $r_0$,
defined by 
\begin{equation}
\left. r^2 F(r) \right\vert_{r=r_0} = 1.65.
\end{equation}
Based on the phenomenological potential models, $r_0$
 corresponds to about $0.5$ fm.
 In this work, however, we use $r_0$ only to set the reference
 scale for comparison with other scheme.
The formula relating $\beta$ and $r_0$ is given
in Ref.\cite{Guagnelli:1998ud},
\begin{equation}
\ln (a/r_0) = -1.6805 - 1.7139(\beta -6) + 0.8155 (\beta -6)^2
         - 0.6667 (\beta -6)^3,
\label{eq:Sommer_a_over_r0}
\end{equation}
which is valid in the region $5.7 \leq \beta \leq 6.57$.
For our lattices, $\beta$ values in Step 7
in Table~\ref{table:parameter-set} are in this range.
In Table~\ref{table:sommer_scale}, we summarize 
$r_0/a$ calculated using Eq.~(\ref{eq:Sommer_a_over_r0}),
and the corresponding $r_0/\tilde{L}_0$.
In Ref.\cite{Guagnelli:1998ud}, the values of $r_0$ in 
the range $5.7 \leq \beta \leq 6.57$ are determined with 
errors linearly
increasing from $0.3\%$ to $0.6\%$.  These errors are
mainly statistical.  
Since we are only aiming at a rough estimate of the absolute scale
in this work, we assign the corresponding 
size of errors to the results in Table~\ref{table:sommer_scale}. 
By extrapolation to the continuum limit,
we obtain\footnote{Here we take the continuum limit and estimate 
the systematic error in the same way as we did in 
subsection \ref{sec:SimDet}.
} 
\begin{equation}
 \frac{r_0}{\tilde{L}_0} = 
 4.75 \pm 0.26.   \hspace{0.5cm}
\label{eq:r0L0}
\end{equation}
Here, the statistical error is negligibly small 
compared to the systematic error.

\begin{table}[h]
\begin{center}
\begin{tabular}{cccc}
\hline
$L_0/a$ & $\beta$ & $r_0/a$ & $r_0/\tilde{L}_0$ \\
\hline
15 & 5.907 & 4.543(17)  & 5.174(19)  \\
18 & 6.000 & 5.368(22)  & 5.096(21) \\ 
21 & 6.087 & 6.196(27)  & 5.041(22)  \\ 
24 & 6.170 & 7.040(33)  & 5.012(23) \\ 
27 & 6.229 & 7.677(37)  & 4.858(23) \\ 
\hline
\end{tabular}
\caption{
The Sommer scale at each $\beta$ of the Step 7,
estimated using Eq.~(\ref{eq:Sommer_a_over_r0}).}
\label{table:sommer_scale}
\end{center}
\end{table}

We can now estimate the $\Lambda$ scale in units of $r_0$.
Since it is obvious from Fig.~\ref{fig:run} 
that $g_w^2$ is well approximated by two-loop perturbative running coupling 
at high energy region, it is reasonable to estimate the scale $\Lambda$ 
by using the value of $g_w^2(1/\tilde{L}_0)$ from the following two-loop 
relation\footnote{From the difference between two-loop and three-loop 
$\beta$ function in the SF scheme, the error in the estimation 
to $\Lambda$ due to the higher order 
effect is about $3\%$.  This is reasonably small compared to other errors
in this work.
}
between $\tilde{L}_0 \Lambda$ and $g_w^2(1/\tilde{L}_0)$,
\begin{equation}
\tilde{L}_0 \Lambda_{\rm WL}^{\rm 2-loop} 
= {\mathrm{e}}^{-\frac{1}{2 b_0 g_w^2(1/\tilde{L}_0) }}
 \left(  \frac{b_0^2 g_w^2(1/\tilde{L}_0)}{b_0 + b_1 g_w^2(1/\tilde{L}_0)} 
 \right)^{-\frac{b_1}{2 b_0^2}}, 
\label{eq:conv}
\end{equation}
where $b_0 = 11/(4\pi)^2$ and $b_1=102/(4\pi)^4$ are the 
one-loop and the two-loop coefficients of the $\beta$ function 
of quenched QCD.
By substituting the value $g_w^2(1/\tilde{L}_0) = 1.728$, we 
find
\begin{equation}
\tilde{L}_0 \Lambda_{\rm WL}^{\rm 2-loop} 
\simeq 0.0399.
\end{equation}
Combining this result with the value of $r_0/\tilde{L}_0$ in 
Eq.~(\ref{eq:r0L0}), we obtain the value of 
$\Lambda_{\rm WL}^{\rm 2-loop}$ in units of $r_0$ as 
\begin{equation}
 r_0 \Lambda_{\rm WL}^{\rm 2-loop} = 
 0.190 \pm 0.010. \hspace{0.5cm}
\end{equation}
We also estimated, in the similar way as above, the value of 
$r_0 \Lambda^{\rm 2-loop}$ in the case of SF scheme by 
using the data reported in Ref.~\cite{Capitani:1998mq} , and found 
the following result: 
\begin{equation}
 r_0 \Lambda_{\rm SF}^{\rm 2-loop} = 0.301 \pm 0.025.
\end{equation}
By fixing $r_0$ as a reference scale, we obtain the following ratio 
of $\Lambda_{\rm SF}^{\rm 2-loop}$ to $\Lambda_{\rm WL}^{\rm 2-loop}$: 
\begin{equation}
 \frac{\Lambda_{\rm SF}^{\rm 2-loop}}{\Lambda_{\rm WL}^{\rm 2-loop}} 
 = 1.58 \pm 0.16.
 \label{eq:ratioWL}
\end{equation}
As a consistency check, we have also
extracted the ground-state potential from
our data at the largest physical
volume {\it via} double exponential fits
~\cite{Fleming:2004hs}, then used the potential to
estimate
$r_0$ and $r_c$ defined as
\begin{equation}
\left. r^2 F(r) \right\vert_{r=r_c} = 0.65.
\end{equation}
Our results on these quantities are well compatible
with those obtained in Ref.~\cite{Necco:2001xg}.

We also estimated the value of 
$\Lambda_{\rm SF}^{\rm 2-loop}/\Lambda_{\rm WL}^{\rm 2-loop}$ 
without relying on the measurement of any low-energy physical quantity. 
This can be achieved by comparing results obtained from the same combinations
of values of $\beta$ and $L_0/a$ in the two schemes,  
with a fixed physical box size $L_0$ which is much smaller than
$1/\Lambda_{{\mathrm{QCD}}}$. 
To this purpose, we have carried out simulations using exactly the
same values of $\beta$ and $L_{0}/a$ as one of the data sets
in Ref.~\cite{Capitani:1998mq} . In Table~\ref{tab:Lambda-comp}, we list the 
values of the coupling constant in the
SF scheme, $g_{\rm SF}^2$ (which is denoted by $\bar{g}^2$ in 
Ref.~\cite{Capitani:1998mq}), and results of $\Lambda_{\rm SF}^{\rm 2-loop}$ 
estimated from them by using Eq.~(\ref{eq:conv}).
\begin{table}[t]
       \begin{center}
       \begin{tabular}{|c|c||c|c||c|c||c|}
       \hline
$L_0/a$ & $\beta$ & $g_{\rm SF}^2$ & $L_0 \Lambda_{\rm SF}^{\rm 2-loop}$ &
$g_w^2$ & $L_0 \Lambda_{\rm WL}^{\rm 2-loop}$ &
$\Lambda_{\rm SF}^{\rm 2-loop}/\Lambda_{\rm WL}^{\rm 2-loop}$      \\
\hline \hline
$10$   & $7.8538$  &$1.8776(93)$  &$0.0539(9)$   & $1.9427(68)$   & $0.06043(71)$ 
&$0.891(18)$  \\
$12$   & $7.9993$  &$1.8811(38)$  &$0.0542(4)$   & $1.8620(66)$   & $0.05232(65)$ 
&$1.036(15)$  \\
$14$   & $8.1380$  &$1.884(11)$    &$0.0545(11)$ & $1.8028(73)$   & $0.04667(68)$ 
&$1.168(29)$  \\
$16$   & $8.2500$  &$1.864(10)$    &$0.0525(10)$ & $1.7662(79)$   & $0.04331(72)$ 
&$1.213(30)$  \\
       \hline
\end{tabular}
       \caption{Values for $g^2$ and $L_0 \Lambda^{\rm 2-loop}$ in the SF and the WL
schemes for 
       several sets of $L_0/a$ and $\beta$. Results of $g_{\rm SF}^2$ are taken from 
       Ref.~\cite{Capitani:1998mq} . The values in parentheses  represent statistical errors 
       in units of the 
last digits.}
       \label{tab:Lambda-comp}
       \end{center}
\end{table}
The values of $g_w^2$ and the corresponding $\Lambda_{\rm WL}^{\rm 2-loop}$ 
are also listed in this table.\footnote{
We notice that the $a/L_0$ dependence of $g_w^2$ is rather large 
partly due to the fact that we used the 
results of the SF-scheme study as inputs to set the scale for each $\beta$.
Therefore, the discretization error could be a 
combination (or the difference ) of those from WL 
scheme and SF scheme. 
This error may be dominated by the error from 
the WL scheme since the effective physical scale 
is the Wilson loop size $R$ but not $L_0$ so that 
one expects $O((a/R)^2)$ error.
} These values result from an
interpolation procedure using the data points shown in Fig.~\ref{fig:g-beta-global}. 
The results for 
$\Lambda_{\rm SF}^{\rm 2-loop}/\Lambda_{\rm WL}^{\rm 2-loop}$  at 
each $L_0/a$ is also listed in the same table.
A fit linear in $a/L_{0}$
to these results
gives the following continuum-limit estimation:
\begin{equation}
\frac{\Lambda_{\rm SF}^{\rm 2-loop}}{\Lambda_{\rm WL}^{\rm 2-loop} }
= 1.78 \pm 0.07 \ ({\rm stat.}) \pm 0.04 \ ({\rm sys.})
\label{eq:ratioSF}
\end{equation}
Here the systematic error was estimated by the difference between values 
in the continuum limit with linear and quadratic extrapolations. 

A comparison between the results of Eqs.~(\ref{eq:ratioWL}) and (\ref{eq:ratioSF}) 
shows the ``universality" of the estimate of 
$\Lambda_{\rm SF}^{\rm 2-loop}/\Lambda_{\rm WL}^{\rm 2-loop}$, 
namely, the two different estimates are consistent with each 
other.\footnote{Three-loop perturbative calculation 
(though quite challenging to carry out it in the WL scheme, while 
already done in the SF scheme), together with the improvement of 
systematic and statistical errors, would enable us to perform a even more precise test 
of this ``universality" by defining the $\Lambda$ 
scale with three-loop $\beta$ function.}
From the theoretical point of view, this ``universality" might be 
a trivial result since quenched QCD has only one scale in the theory. However, from the
numerical point of view, it is a rather non-trivial consistency check 
since one estimation involves the measurement of a low-energy physical 
quantity while the other is completely from high-energy physics.

We notice that the error of $g_w^2$ is large 
partly due to the fact that we used the 
SF as inputs to set the scale for each $\beta$.
Therefore, the discretization error could be a 
combination (or the difference ) of those from WL 
scheme and SF scheme. 
This error may be dominated by the error from 
the WL-scheme since the effective physical scale 
is the Wilson loop size $R$ but not $L_0$ so that 
one expects $O((a/R)^2)$ error.

\section{Summary}

We proposed a new scheme for the determination of the running coupling 
on the lattice. Our method is based on the measurement of the finite volume 
dependence of the Wilson loop. Unlike the SF scheme, our method does 
not have any $O(a)$ discretization error, therefore the systematic 
effect arising 
from the extrapolation to the continuum limit is expected to be quite small. 
We showed results of numerical study for the quenched QCD 
as a feasibility test of our scheme.  These results confirmed that our
method led to
the step-scaling of the coupling which is consistent 
with the perturbative running coupling at high energy. We also showed that 
the coupling calculated by this newly proposed scheme deviates from 
the two-loop approximation below a certain energy scale. 
This deviation arises from the effects that are not captured by 
the two-loop approximation. We have confirmed that our scheme  
works well for the calculation of the running coupling with relatively 
small number of gauge configurations, demonstrating that the statistical error 
is under control by properly choosing the smearing level and $r$. 
We expect that this new method is also applicable to the calculation of the
running 
coupling in other gauge theories, including the $SU(N)$ 
gauge theory with a large number of dynamical fermions, which will be 
studied in our future work.

\section*{Acknowledgements}

We thank Biagio Lucini and Antonio Rago for helpful
discussions. We also thank Koichi Yamawaki for organizing 
a workshop on related topics at Nagoya University, which 
stimulated our work.
This work is supported in part by the Grant-in-Aid of the Ministry
of Education (Nos. 19540286, 19740160, 20039005, 20105002 and 20740133).
E.~B. is supported by the FWF Doktoratskolleg Hadrons in Vacuum,
Nuclei and Stars (DK W1203-N08).
A.~F. acknowledges the support of JSPS, Grant N.19GS0210.
M.~K. is supported by Los Alamos National Laboratory 
under DE-AC52-06NA25396.
C.-J D. L. is supported by the National Science Council of Taiwan {\it via}
grant number 96-2112-M-009-020-MY3.
During this work, T.Y. has been the Yukawa Fellow supported by
Yukawa Memorial Foundation.
Numerical simulation was carried out on the vector supercomputer
NEC SX-8 at YITP, Kyoto University, and SX-8 at RCNP, Osaka University.
We are grateful to YITP, Kyoto and NCTS, Hsinchu for hospitality
during the progress of this work.

\appendix

\section{Formulas (\ref{resa2}), and (\ref{resa3})}
\label{app:summation}
In the following, we will briefly show how to obtain formulas (\ref{resa2}), 
and (\ref{resa3}). Basically the method we use is a repetition of the generic
technique we have adopted in Sec. \ref{creutzratio}: we first perform the sum
over $m$, and then use the Chowla-Selberg formula, which rearranges the expression
as a sum of some analytic function plus a series, suppressed by the presence of the
Kelvin functions $K_\nu(x)$. 
For the first term one has:
\bea
&&\sum_{m=-\infty}^{+\infty}\sum_{p,q=1}^{\infty} 
{1\over \left({p^2}+{q^2} +|m+R/L_0|^2\right)^{s} }\nonumber\\
&=&\sum_{p,q=1}^{\infty} 
\left[
\frac{\sqrt{\pi } \left(p^2+q^2\right)^{\frac{1}{2}-s} \Gamma
\left(s-\frac{1}{2}\right)}{\Gamma (s)}\right.\nonumber\\
&+& \left. \frac{4 \pi ^s}{\Gamma (s)}
\left(p^2+q^2\right)^{\frac{1}{4}-\frac{s}{2}} \sum _{j=1}^{\infty }
j^{s-\frac{1}{2}} \cos (2 j \pi  R/L_0) K_{s-\frac{1}{2}}\left(2 j \pi 
 \sqrt{p^2+q^2}\right)\right]\nonumber\\
&=&\sum_{p,q=1}^{\infty} \left[i
\frac{\sqrt{\pi } \left(p^2+q^2\right)^{\frac{1}{2}-s} \Gamma
\left(s-\frac{1}{2}\right)}{\Gamma (s)}\right.
+\left.4 \sum _{j=1}^{\infty } \cos (2 j \pi  R/L_0) K_0\left(2 j \pi 
\sqrt{p^2+q^2}\right)\right]~.
\label{dc}
\eea
In the above formula, the first term disappears upon derivation with respect to $R$
and thus will not contribute to $k$. The second term is exponentially suppressed due
to the presence of $K_0(z)$ and, thus, regular. An analogous procedure applies to
$A_3(R/L_0)$:
\bea
&&\sum_{m=-\infty}^{+\infty}\sum_{q=1}^{\infty}
{ 1\over \left({q^2}+|m+R/L_0|^2\right)^{s}}
\nonumber\\
&=& \sum_{q=1}^{\infty}\frac{\sqrt{\pi }
 \Gamma (2s-1)}{\Gamma \left(\frac{1}{2}\right)}+4 \sum _{q,j=1}^{\infty } \cos (2
j \pi R/L_0) K_0\left(2 j \pi q\right)~.
\label{nc}
\eea
As before, the first term is independent of $R$ and disappears, when differentiated
with respect to $R$. 
The second term is exponentially suppressed and regular.

\section{Summary of Simulation Parameters and Numerical Results}
\label{app:simulation_parameters}
Here, we summarize simulation parameters, and numerical results 
obtained from those simulations. In the tables below, $N_{\rm conf}$, 
$N_{\rm sweep}$ and $N_{\rm bin\, size}$ respectively represent 
number of gauge configurations from which measurement were taken, 
number of sweeps between each configurations, and number of 
bin size when we estimate the statistical error. The values in parentheses  
in the column of $\tilde{g}_w^2$ represent statistical errors in units of the 
last digits. We also listed the values of $\tilde{g}_w^2$ in the continuum 
limit with the magnitude of total error at each step.

\begin{center}
\begin{tabular}{|c|c|c|c|c|c|}
       \hline
\multicolumn{6}{|c|}{Step $1$} \\
\hline \hline
$L_0/a$ & $\beta$ & $N_{\rm conf}$ & $N_{\rm sweep}$ & $N_{\rm bin\, size}$ &
$\tilde{g}_w^2$ \\
\hline
$15$ & $8.31$ & $200$ & $1000$ & $1$ & $0.2846(24)$  \\
$18$ & $8.25$ & $200$ & $1000$ & $1$ & $0.2999(24)$  \\
$21$ & $8.27$ & $400$ & $1000$ & $1$ & $0.3100(21)$  \\
$24$ & $8.32$ & $400$ & $1000$ & $1$ & $0.3119(27)$  \\
$27$ & $8.40$ & $200$ & $1000$ & $1$ & $0.3137(51)$  \\
\hline
\hline
\multicolumn{6}{|c|}{continuum limit :\ \ $\tilde{g}_w^2 = 0.328 \pm 0.010 $} \\
\hline
\end{tabular}
\begin{tabular}{|c|c|c|c|c|c|}
       \hline
\multicolumn{6}{|c|}{Step $2$} \\
\hline \hline
$L_0/a$ & $\beta$ & $N_{\rm conf}$ & $N_{\rm sweep}$ & $N_{\rm bin\, size}$ &
$\tilde{g}_w^2$ \\
\hline
$15$ & $7.80$ & $200$ & $1000$ & $1$ & $0.3371(26)$  \\
$18$ & $7.83$ & $200$ & $1000$ & $1$ & $0.3465(29)$  \\
$21$ & $7.86$ & $200$ & $1000$ & $1$ & $0.3482(36)$  \\
$24$ & $7.91$ & $200$ & $1000$ & $1$ & $0.3585(48)$  \\
$27$ & $7.97$ & $200$ & $1000$ & $1$ & $0.3632(66)$  \\
       \hline
\hline
\multicolumn{6}{|c|}{continuum limit :\ \ $\tilde{g}_w^2 = 0.371 \pm 0.012 $} \\
\hline
\end{tabular}
\begin{tabular}{|c|c|c|c|c|c|}
       \hline
\multicolumn{6}{|c|}{Step $3$} \\
\hline \hline
$L_0/a$ & $\beta$ & $N_{\rm conf}$ & $N_{\rm sweep}$ & $N_{\rm bin\, size}$ &
$\tilde{g}_w^2$ \\
\hline
$15$ & $7.44$ & $200$ & $1000$ & $1$ & $0.3831(31)$  \\
$18$ & $7.45$ & $400$ & $1000$ & $1$ & $0.4056(25)$  \\
$21$ & $7.49$ & $400$ & $1000$ & $1$ & $0.4125(33)$  \\
$24$ & $7.55$ & $400$ & $1000$ & $1$ & $0.4278(50)$  \\
$27$ & $7.61$ & $200$ & $1000$ & $1$ & $0.4249(92)$  \\
       \hline
       \hline
\multicolumn{6}{|c|}{continuum limit :\ \ $\tilde{g}_w^2 = 0.445 \pm 0.009 $} \\
\hline
\end{tabular}
\begin{tabular}{|c|c|c|c|c|c|}
       \hline
\multicolumn{6}{|c|}{Step $4$} \\
\hline \hline
$L_0/a$ & $\beta$ & $N_{\rm conf}$ & $N_{\rm sweep}$ & $N_{\rm bin\, size}$ &
$\tilde{g}_w^2$ \\
\hline
$15$ & $6.968$ & $10000$ & $1$ & $100$ & $0.4829(9)$  \\
$18$ & $7.040$ & $10000$ & $1$ & $100$ & $0.4959(12)$  \\
$21$ & $7.076$ & $10000$ & $1$ & $100$ & $0.5146(15)$  \\
$24$ & $7.156$ & $10000$ & $1$ & $100$ & $0.5204(20)$  \\
$27$ & $7.243$ & $10000$ & $1$ & $100$ & $0.5181(24)$  \\
       \hline
       \hline
\multicolumn{6}{|c|}{continuum limit :\ \ $\tilde{g}_w^2 = 0.547 \pm 0.004 $} \\
\hline

\end{tabular}
\begin{tabular}{|c|c|c|c|c|c|}
       \hline
\multicolumn{6}{|c|}{Step $5$} \\
\hline \hline
$L_0/a$ & $\beta$ & $N_{\rm conf}$ & $N_{\rm sweep}$ & $N_{\rm bin\, size}$ &
$\tilde{g}_w^2$ \\
\hline
$15$ & $6.571$ & $10000$ & $1$ & $100$ & $0.6307(14)$  \\
$18$ & $6.656$ & $10000$ & $1$ & $100$ & $0.6489(19)$  \\
$21$ & $6.734$ & $10000$ & $1$ & $100$ & $0.6606(27)$  \\
$24$ & $6.797$ & $10000$ & $1$ & $100$ & $0.6794(33)$  \\
$27$ & $6.871$ & $10000$ & $1$ & $100$ & $0.6931(41)$  \\
       \hline
       \hline
\multicolumn{6}{|c|}{continuum limit :\ \ $\tilde{g}_w^2 = 0.719 \pm 0.024 $} \\
\hline
\end{tabular}
\begin{tabular}{|c|c|c|c|c|c|}
       \hline
\multicolumn{6}{|c|}{Step $6$} \\
\hline \hline
$L_0/a$ & $\beta$ & $N_{\rm conf}$ & $N_{\rm sweep}$ & $N_{\rm bin\, size}$ &
$\tilde{g}_w^2$ \\
\hline
$15$ & $6.207$ & $10000$ & $1$ & $100$ & $0.978(3)$  \\
$18$ & $6.303$ & $10000$ & $1$ & $100$ & $1.016(5)$  \\
$21$ & $6.377$ & $10000$ & $1$ & $100$ & $1.069(6)$  \\
$24$ & $6.463$ & $10000$ & $1$ & $100$ & $1.075(7)$  \\
$27$ & $6.546$ & $10000$ & $1$ & $100$ & $1.074(9)$  \\
       \hline
       \hline
\multicolumn{6}{|c|}{continuum limit :\ \ $\tilde{g}_w^2 = 1.144 \pm 0.037 $} \\
\hline
\end{tabular}
\begin{tabular}{|c|c|c|c|c|c|}
       \hline
\multicolumn{6}{|c|}{Step $7$} \\
\hline \hline
$L_0/a$ & $\beta$ & $N_{\rm conf}$ & $N_{\rm sweep}$ & $N_{\rm bin\, size}$ &
$\tilde{g}_w^2$ \\
\hline
$15$ & $5.907$ & $10000$ & $1$ & $100$ & $1.811(5)$  \\
$18$ & $6.000$ & $10000$ & $1$ & $100$ & $1.833(7)$  \\
$21$ & $6.087$ & $10000$ & $1$ & $100$ & $1.846(12)$  \\
$24$ & $6.170$ & $10000$ & $1$ & $100$ & $1.861(13)$  \\
$27$ & $6.229$ & $10000$ & $1$ & $100$ & $1.917(27)$  \\
       \hline
\hline
\multicolumn{6}{|c|}{continuum limit :\ \ $\tilde{g}_w^2 = 1.914 \pm 0.042 $} \\
\hline
\end{tabular}
\end{center}

\addcontentsline{toc}{chapter}{Bibliography} 
\bibliographystyle{prsty} 
\bibliography{refs}

\end{document}